\documentclass{emulateapj}
\usepackage{color}
\usepackage{comment}

\usepackage{natbib,aas_macros}
\newcommand{\Lya}{Ly$\alpha$}
\newcommand{\OII}{[O\,{\sc ii}]}
\newcommand{\Ha}{H$\alpha$}
\newcommand{\Hb}{H$\beta$}

\newcommand{\CIV}{C{\sc iv}}

\newcommand{\NII}{[N\,{\sc ii}]}
\newcommand{\OIII}{[O\,{\sc iii}]}
\newcommand{\SII}{[S\,{\sc ii}]}
\newcommand{\xmm}{{\it XMM-Newton}}
\newcommand{\fesc}{$f_{\rm esc}^{\rm ion}$}
\newcommand{\fescLya}{$f_{\rm esc}^{{\rm Ly}\alpha}$}
\newcommand{\ergs}{erg\,s$^{-1}$}
\newcommand{\ergscm}{erg\,s$^{-1}$\,cm$^{-2}$}

\newcommand{\Oabundance}{$12+\log ({\rm O/H})$}
\newcommand{\HII}{H{\sc ii}}

\slugcomment{Accepted for publication in ApJ}

\shorttitle{
High Ionization State and Low Oxygen Abundance in LAEs
}
\shortauthors{Nakajima et al.}

\begin{document}

\title{
First Spectroscopic Evidence for High Ionization State and \\
Low Oxygen Abundance in Ly$\alpha$ Emitters
\altaffilmark{\dag}
\altaffilmark{\ddag}
}

\author{Kimihiko Nakajima  \altaffilmark{1,3},
       Masami Ouchi        \altaffilmark{2,3},
       Kazuhiro Shimasaku  \altaffilmark{1,4},
       Takuya Hashimoto    \altaffilmark{1},
       Yoshiaki Ono        \altaffilmark{1,2}, \\ and 
       Janice C. Lee       \altaffilmark{5,6,7,8}
      }

\email{nakajima@astron.s.u-tokyo.ac.jp}

\altaffiltext{1}{%
Department of Astronomy, Graduate School of Science,
The University of Tokyo, 7-3-1 Hongo, Bunkyo-ku, Tokyo 113-0033,
Japan
}
\altaffiltext{2}{%
Institute for Cosmic Ray Research, The University of Tokyo,
5-1-5 Kashiwanoha, Kashiwa, Chiba 277-8582, Japan
}
\altaffiltext{3}{%
Kavli Institute for the Physics and Mathematics of the Universe (WPI),
The University of Tokyo, 5-1-5 Kashiwanoha, Kashiwa, 
Chiba 277-8583, Japan
}
\altaffiltext{4}{%
Research Center for the Early Universe, Graduate School of Science,
The University of Tokyo, Tokyo 113-0033, Japan
}
\altaffiltext{5}{%
Space Telescope Science Institute, Baltimore, MD, USA
}
\altaffiltext{6}{%
Visiting Astronomer, SSC/IPAC, Caltech, Pasadena, CA, USA
}
\altaffiltext{7}{%
Observatories of the Carnegie Institution of Washington,
813 Santa Barbara Street, Pasadena, CA 91101, USA
}
\altaffiltext{8}{%
Carnegie Fellow
}

\altaffiltext{\dag}{%
Some of the data presented herein were obtained at the W.M. Keck Observatory, 
which is operated as a scientific partnership among the California Institute 
of Technology, the University of California and the National Aeronautics and 
Space Administration. The Observatory was made possible by the generous 
financial support of the W.M. Keck Foundation.
}

\altaffiltext{\ddag}{%
Based in part on data collected at Subaru Telescope,
which is operated by the National Astronomical Observatory of Japan.
}

\begin{abstract}

We present results from Keck/NIRSPEC and Magellan/MMIRS follow-up 
spectroscopy of Ly$\alpha$ emitters (LAEs) at $z=2.2$ identified in our 
Subaru narrowband survey. We successfully detect \Ha\ emission from seven 
LAEs, and perform a detailed analysis of six LAEs free from AGN activity, 
two out of which, CDFS-3865 and COSMOS-30679, have \OII\ and \OIII\ line 
detections. They are the first \OII-detected LAEs at high-$z$, and their 
\OIII/\OII\ ratios and $R23$-indices provide the first simultaneous 
determinations of ionization parameter and oxygen abundance for LAEs.
CDFS-3865 has a very high ionization parameter
($q_{ion}=2.5^{+1.7}_{-0.8} \times 10^8$\,cm\,s$^{-1}$) and a low oxygen
abundance ($12+\log ({\rm O/H})=7.84^{+0.24}_{-0.25}$) in contrast with 
moderate values of other high-$z$ galaxies such as Lyman-break galaxies 
(LBGs). COSMOS-30679 also possesses a relatively high ionization parameter
($q_{ion}=8^{+10}_{-4} \times 10^7$\,cm\,s$^{-1}$) and a low oxygen abundance
($12+\log ({\rm O/H})=8.18^{+0.28}_{-0.28}$). Both LAEs appear to fall below 
the mass-metallicity relation of $z\sim 2$ LBGs. Similarly, a low 
metallicity of $12+\log ({\rm O/H})<8.4$ is independently indicated for 
typical LAEs from a composite spectrum and the \NII/\Ha\ index. Such high 
ionization parameters and low oxygen abundances can be found in local 
star-forming galaxies, but this extreme local population occupies only 
$\sim 0.06$\% of the SDSS spectroscopic galaxy sample with a number density
$\sim 100$ times smaller than that of LAEs. With their high ionization 
parameters and low oxygen abundances, LAEs would represent an early stage 
of galaxy formation dominated by massive stars in compact star-forming 
regions. High-$q_{ion}$ galaxies like LAEs would produce ionizing photons 
efficiently with a high escape fraction achieved by density-bounded \HII\ 
regions, which would significantly contribute to cosmic reionization at 
$z>6$.

\end{abstract}


\section{INTRODUCTION} \label{sec:introduction}

\Lya\ emitters (LAEs), galaxies commonly observed at high redshifts with
strong \Lya\ emission, are considered to be low-mass, young galaxies
as suggested from their small sizes, faint continua, and low masses
inferred from spectral energy distribution (SED) fitting
(e.g., 
\citealt{venemans2005,gawiser2006,pirzkal2007,overzier2008,malhotra2012}).
LAEs are therefore likely to represent galaxies in the early stages of 
galaxy evolution.
More directly, \citet{cowie2011} investigate rest-frame optical nebular 
lines of LAEs at low redshifts ($z\sim0.3$). 
The authors find that a large portion of LAEs ($75$\,\%) have equivalent 
width (EW) of \Ha\ $>100$\,\AA, and that LAEs on average have lower 
metallicities and younger ages than the UV-continuum sample. These findings 
are consistent with the idea that LAEs are galaxies in early stages 
of galaxy formation.
At higher redshifts, LAEs are efficiently detected thanks to narrowband 
imaging techniques which have enriched our knowledge about the younger 
universe
(e.g., \citealt{CH1998,MR2002,ouchi2003,gawiser2006,shimasaku2006,%
gawiser2007,gronwall2007,ouchi2008,nilsson2009,guaita2010,finkelstein2011,%
nakajima2012}). 
However, more direct evidence supporting the idea that high redshift LAEs
are in an early evolutionary phase of formation is still needed.

The gas-phase metallicity is a key property of galaxies, since it is a 
record of their star-formation histories. This physical quantity is 
relatively easily constrained with line ratios of nebular lines at 
rest-frame optical wavelengths (e.g., \citealt{pagel1979,KD2002}, and 
references therein). Another key quantity is the ionization parameter, 
defined as the ratio of the mean ionizing photon flux to the mean 
hydrogen atom density. Since the excitation of the \HII\ region is 
sensitive to the age distribution of the exciting stars, the ionization 
parameter provides a rough estimate of age of a galaxy (e.g., 
\citealt{dopita2006}). A large ionization parameter is observed from a 
galaxy dominated with massive stars, which is a sign that the galaxy is 
in an early stage of galaxy formation. In addition, the ionization 
parameter depends on the optical depth in an \HII\ region (e.g., 
\citealt{brinchmann2008}); a large ionization parameter may be instead 
due to a low optical depth and a high escape fraction of ionizing photons. 
Therefore, constraining ionization parameters for LAEs may provide an 
independent clue to whether or not LAEs, commonly observed star-forming 
galaxies at high redshift, significantly contribute to the cosmic 
reionization in the early universe. Determining the ionization parameter 
requires two emission lines of different ionization stages of a same 
element, such as \OII\ and \OIII\ (e.g., \citealt{KD2002}).

However, well-defined samples of LAEs are generally located at very high 
redshifts ($3<z<7$), where rest-frame optical nebular lines are redshifted 
into infrared wavelengths. Recently, \citet{finkelstein2011} and  
\citet{nakajima2012} instead take advantage of LAEs at moderately high 
redshifts ($z\sim 2$), where the nebular lines up to \Ha\ are observable 
from the ground in the near-infrared (NIR) windows.

These studies extend the metallicity censuses to galaxies with strong 
\Lya\ emission. \citet{finkelstein2011} obtain NIR spectra for two 
$z\sim 2.4$ LAEs, and find that at least one LAE appears to be less 
chemically enriched than $z\sim 2$ continuum-selected galaxies at similar 
stellar masses \citep{erb2006a}. \citet{nakajima2012} estimate average 
line fluxes for \OII$\lambda 3727$ and \Ha+\NII$\lambda6584$ for $z=2.2$ 
LAEs by stacking $1.18$ and $2.09\,\mu$m narrowband images for more than 
$100$ LAEs, and place a firm lower-limit for the average metallicity of 
this population. Interestingly, the lower-limit is higher than expected 
for its stellar mass at the redshift.

In contrast to the gas-phase metallicities, the ionization parameters of 
LAEs are unknown. Such measurements have only been obtained for bright 
galaxies such as Lyman-break galaxies (LBGs) or strongly lensed galaxies 
at high-$z$ (e.g., \citealt{pettini2001,hainline2009,richard2011}).
Although the current number of measurements is small, ionization 
parameters of such high-$z$ galaxies are likely to be higher on average 
than those of local galaxies. A comparison of ionization parameters for 
LBGs and LAEs will enable us to discuss whether or not LAEs are young 
galaxies at high redshifts.

In order to obtain reliable measurements of properties including metallicity 
and ionization parameter of LAEs at high redshifts, we have carried out a 
large survey for $z=2.2$ LAEs, using our custom narrowband filter NB387 with 
Subaru/Suprime-Cam.
At $z=2.2$, important nebular lines such as \OII$\lambda 3727$, \Hb, 
\OIII$\lambda\lambda 5007,4959$, \Ha, \NII$\lambda 6584$, are observable from 
the ground. Initial results were based on using three narrowbands to detect 
\Lya, \OII, and \Ha\ over the same volumes (\citealt{nakajima2012}; see also 
\citealt{lee2012} and \citealt{ly2011} for the two NIR narrowbands and the 
New\Ha\ Survey).

In this paper, we present results from NIR spectroscopy. We used Keck/NIRSPEC 
and Magellan/MMIRS spectrographs, and successfully detected \Ha\ emission 
from seven LAEs. The number is double the previous number of high-$z$ LAEs 
with NIR spectra (two from \citealt{finkelstein2011} and two from 
\citealt{mclinden2011}), and allows us to begin to examine statistical 
variation of rest-frame optical spectroscopic properties of LAEs at $z\sim 2$.
Our first NIR spectroscopic result discusses the kinematics of LAEs and is 
presented in \citet{hashimoto2013}. As a companion study, this paper presents 
the ionization and chemical properties of LAEs based on multiple nebular lines.
Remarkably, we detected \OII\ and \OIII\ lines from two LAEs , which provide 
ionization parameter estimates for LAEs for the first time. In addition, the 
oxygen lines allow us to determine oxygen abundances that compliment the 
previous spectroscopic constraints on metallicity of LAEs from the \NII/\Ha\ 
index.

We also investigate \Lya\ and \Ha\ hydrogen lines for LAEs. A comparison of 
observed \Lya/\Ha\ ratios with the Case B recombination value (\Lya/\Ha\ 
$=8.7$; \citealt{brocklehurst1971}) provides important insights into the 
physical mechanisms causing the strong \Lya\ emission in LAEs. Furthermore, 
we study \Lya\ and \Ha\ equivalent widths which probe star-formation history, 
stellar age, and metallicity.

This paper is organized as follows. We describe the data in \S\ref{sec:data}. 
The detection and measurement of emission lines in the NIR spectroscopy is 
summarized in \S\ref{sec:line_measure}. In \S\ref{sec:properties}, we derive 
properties of LAEs including estimates of ionization parameter, metallicity, 
and SFR from the rest-frame optical nebular lines. We also check for the 
presence of active galactic nuclei (AGN) in the LAEs. In 
\S\ref{sec:discussion}, we compare LAEs with other galaxies in terms of 
their ionization state, metallicity, and SFR. We then discuss the 
implications. We also discuss the physical properties inferred from \Lya\ 
and \Ha\ emission. We conclude the paper in \S\ref{sec:summary} with a 
summary. Throughout this paper, magnitudes are given in the AB system 
\citep{oke1974}, and we assume a standard $\Lambda$CDM cosmology with 
$(\Omega_m,\Omega_{\Lambda},H_0)=
(0.3,0.7,70\,{\rm km}\,{\rm s}^{-1}\,{\rm Mpc}^{-1})$.

\section{NIR spectroscopic data} \label{sec:data}

\subsection{Sample Construction} \label{ssec:data_NB387}

\begin{deluxetable*}{lccccccccccc}
\tablecolumns{12}
\tabletypesize{\scriptsize}
\tablecaption{Summary of the LAE sample with NIR spectroscopy%
\label{tbl:sum_data}}
\tablehead{%
\colhead{Object} &
\colhead{R.A.} &
\colhead{Decl.} &
\colhead{NB387} &
\colhead{$U-$NB387} &
\colhead{$B-$NB387} &
\colhead{EW(\Lya)} & 
\colhead{$F$(\Lya)} &
\multicolumn{3}{c}{EXPTIME\tablenotemark{(6)}} &
\colhead{Instr.} \\
\colhead{} &
\colhead{\tiny (1)} &
\colhead{\tiny (1)} &
\colhead{\tiny (2)} &
\colhead{\tiny (3)} &
\colhead{\tiny (3)} &
\colhead{\tiny (4)} &
\colhead{\tiny (5)} &
\colhead{($J$)} &
\colhead{($H$)} &
\colhead{($K$)} &
\colhead{\tiny (7)} 
}
\startdata
COSMOS-08501  & 
10:01:16.80 & 
$+$02:05:36.26 & 
$23.94$ &
$1.41\pm 0.09$ &
$2.05\pm 0.10$ &
$255\pm 26$ & 
$24.0\pm 1.1$ &
\nodata & \nodata & $3600$ &
N \\
COSMOS-13636  & 
09:59:59.38 & 
$+$02:08:38.36 & 
$23.53$ &
$0.91\pm 0.05$ &
$1.03\pm 0.06$ &
$73\pm 5$ & 
$32.4\pm 1.4$ &
\nodata & \nodata & $5400$ &
N \\
COSMOS-30679  & 
10:00:29.81 & 
$+$02:18:49.00 & 
$23.63$ &
$0.86\pm 0.05$ &
$0.55\pm 0.06$ &
$34\pm 3$ & 
$19.1\pm 1.4$ &
$5400$  & $7200$  & $6300$ &
N \\
COSMOS-30679\tablenotemark{(\dag)}  & 
\nodata & 
\nodata & 
$23.73$ &
$1.20\pm 0.07$ &
$1.16\pm 0.07$ &
$87\pm 7$ & 
$23.2\pm 1.8$ &
\nodata & \nodata & \nodata &
\nodata \\
COSMOS-43982  & 
09:59:54.39 & 
$+$02:26:29.96 & 
$23.83$ &
$1.14\pm 0.08$ &
$1.30\pm 0.07$ &
$105\pm 8$ & 
$30.3\pm 1.3$ &
\nodata & \nodata & $3600$ &
N \\
HHDFN-18325   & 
12:36:23.36 & 
$+$62:06:05.10 & 
$21.84$ &
$1.26\pm 0.02$ &
$1.50\pm 0.02$ &
$122\pm 2$ & 
$125.\pm 1.2$ &
 \nodata & \nodata & $3600$ &
N \\
HHDFN-18431   & 
12:36:25.62 & 
$+$62:05:37.43 & 
$23.20$ &
$1.15\pm 0.03$ &
$1.71\pm 0.04$ &
$156\pm 6$ & 
$37.8\pm 0.7$ &
 \nodata & \nodata & $3600$ &
N \\
\\
CDFS-3865    & 
03:32:32.31 & 
$-$28:00:52.20 & 
$22.29$ &
$1.42\pm 1.23$ &
$1.01\pm 0.42$ &
$64\pm 29$ & 
$84.0\pm 13.9$ &
$5100$  & \multicolumn{2}{c}{10800} &
N/M \\
CDFS-6482    & 
03:32:49.34 & 
$-$27:59:52.35 & 
$23.26$ &
$1.47\pm 1.92$ &
$0.96\pm 0.66$ &
$75\pm 52$ & 
$41.9\pm 22.0$ &
\nodata & \multicolumn{2}{c}{10800} &
M \\
SSA22-8043   & 
22:17:47.33 & 
$+$00:08:28.36 & 
$24.51$ &
$0.66\pm 0.06$ &
$0.52\pm 0.06$ &
$28\pm 4$ & 
$6.3\pm 1.1$ &
\nodata & \multicolumn{2}{c}{10800} &
M  
\enddata
\tablecomments{
(1) Coordinates are in J2000.
(2) NB387 aperture magnitude. The diameter of the aperture is $2\farcs 0$,
and the typical error is $0.04$\,mag.
(3) $U-$NB387 and $B-$NB387 colors and their $1\sigma$ errors 
calculated from aperture magnitudes.  
(4) Rest-frame EW of \Lya\ emission line in units of \AA\ calculated 
from $B-$NB387 color and redshift of \Ha. For HHDFN objects whose \Ha\ 
are not detected, we assume $z=2.18$, which corresponds to the wavelength 
of the peak of the NB387 transmission curve.
(5) Flux of \Lya\ emission line in units of $10^{-17}$\,\ergscm\ 
calculated from EW(\Lya) and $B$ band total magnitude. 
(6) Exposure time with NIRSPEC $J$, $H$, and $K$ bands in units of second. 
For the CDFS and SSA22 objects, their $H$ and $K$ bands spectra were taken 
with MMIRS using the $HK$ grism.
(7) Instrument used for the spectroscopy. ``N" stands for NIRSPEC, and 
``M" for MMIRS. For CDFS-3865, its $J$ band spectrum was obtained with 
NIRSPEC, and $H$ and $K$ bands spectra with MMIRS.
(\dag) Values obtained after removing the contribution from an 
adjacent object (\S\ref{ssec:SEDfit}).
}
\end{deluxetable*}

We carried out NB387 ($\lambda_c=3870$\,\AA\ and FWHM $=94$\,\AA)
imaging observation with Subaru/Suprime-Cam 
\citep{miyazaki2002} on 2009 December 14-16 and 19-20 to search for 
$z=2.2$ LAEs. A total of $\sim 1.5$ square degrees are covered, with 
pointings in the following five fields: 
the Subaru/\xmm\ Deep Survey (SXDS) field \citep{furusawa2008}, 
the COSMOS field \citep{scoville2007}, 
the Chandra Deep field South (CDFS; \citealt{giacconi2001}),
the Hawaii Hubble Deep Field North (HHDFN; \citealt{capak2004}), 
and the SSA22 field (e.g., \citealt{steidel2000}).
The first results of the NB387 survey in the SXDS field have been 
presented in \citet{nakajima2012}. For the other fields, we select $z=2.2$ 
LAE candidates in the same manner as presented in \citet{nakajima2012}, 
and a summary of the full $\sim 1.5$ square degrees survey will be 
presented elsewhere (K. Nakajima et al. in preparation). Briefly, objects 
which are bright in NB387 compared to the $U$ and $B$ bands are selected 
as LAE candidates. The color criterion for SXDS, COSMOS, and SSA22 is 
$u^{\star}-\rm{NB387}>0.5 \,\&\&\, B-\rm{NB387}>0.2$, while 
$U-\rm{NB387}>0.8 \,\&\&\, B-\rm{NB387}>0.2$ for CDFS and 
$U-\rm{NB387}>0.5 \,\&\&\, B-\rm{NB387}>0.2$ for HHDFN. These criteria 
results in selecting objects with emission line possessing 
EW$_{\rm rest}\gtrsim 30$\,\AA. The $2\sigma$ photometric errors in 
$u^{\star}-$NB387 (or $U-$NB387) are additionally used as the minimum excess 
for an object to enter the sample. Note that this narrowband excess 
criterion is only significant for CDFS, where broadband data are shallow 
with respect to the NB387 data. Although a much deeper U-band data exist 
in CDFS \citep{nonino2009}, we use the shallower U-band since it has a 
wider coverage%
\footnote{
In addition, CDFS-3865 is located near the edge of the deeper 
image. 
}. Since the U-band data is not directly used when deriving physical 
properties, the shallowness is not significant in this paper. Interlopers 
such as \OII\ emitters, \CIV\ emitters, or active galactic nuclei (AGNs) 
are removed from the catalog by using UV, X-ray and/or radio data.

Seven candidates were selected for NIRSPEC follow-up. We selected 
COSMOS-13636, COSMOS-30679, and COSMOS-43982, since their \Lya\ spectra 
had been obtained with Magellan/MagE (M. Rauch et al. in preparation). 
COSMOS-08501, HHDFN-18325, and HHDFN-18431 were selected because of their 
large \Lya\ equivalent widths and fluxes measured from the NB387 imaging 
data. CDFS-3865, whose rest-frame optical spectrum has in part already 
been taken with MMIRS, was also observed with NIRSPEC for its \OII\ 
detection. Table \ref{tbl:sum_data} summarizes the details of the sample.

\begin{figure*}
\epsscale{1.00}
\plotone{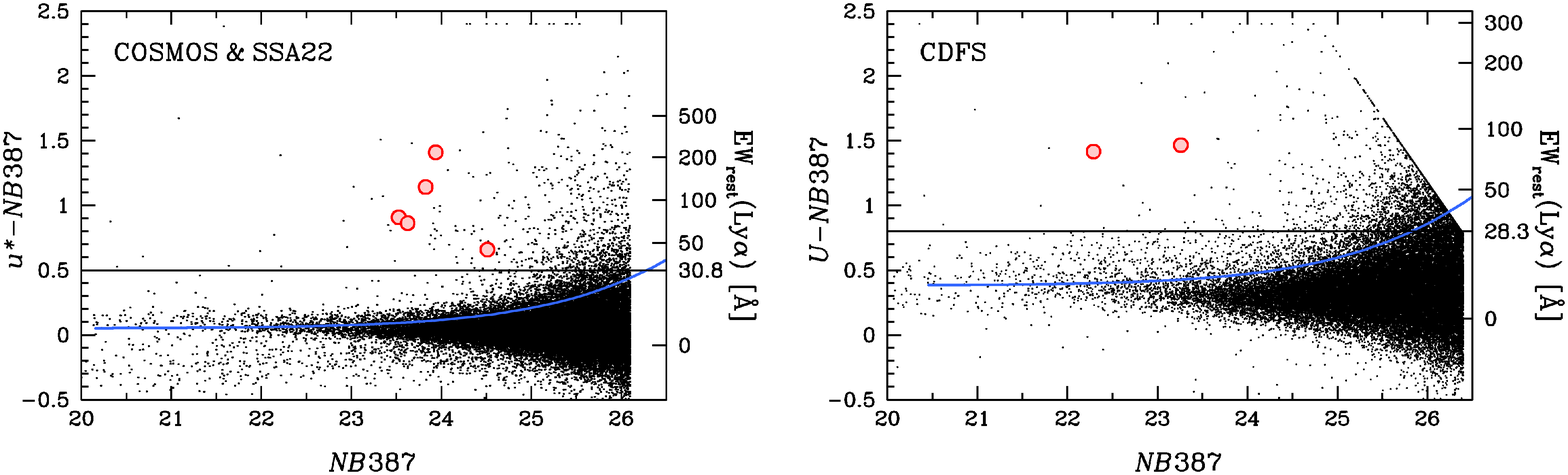}
\caption{
Distribution of all objects in the $u^*-$NB387 vs. NB387 plane detected 
in the COSMOS and SSA22 fields (left) and in the $U-$NB387 vs. NB387 plane 
in the CDFS field (right). The red circles show the LAEs presented in this 
paper, and the black dots show the NB387-detected objects. For the purpose 
of display, objects whose $u^*-$NB387 (or $U-$NB387) colors exceed $2.4$ are 
plotted at $u^*-$NB387 ($U-$NB387) $=2.4$. The horizontal solid line in 
each panel shows the selection threshold of $u^*-$NB387 ($U-$NB387) and 
the blue curve indicates the $2\sigma$ photometric error in $u^*-$NB387 
($U-$NB387) for objects with $u^*-$NB387=0.05 ($U-$NB387=0.38) in the COSMOS 
and SSA22 (CDFS) fields, which is the average color of all the objects. 
The right $y$-axis in each panel shows the rest-frame \Lya\ equivalent 
width of $z=2.18$ LAEs with the $u^*-$NB387 ($U-$NB387) color corresponding 
to the left $y$-axis.
\label{fig:CM}}
\end{figure*}

For the MMIRS observation, we observed CDFS and SSA22 fields with one mask 
each. Details of the observation and data reduction procedures are 
presented in \citet{hashimoto2013}. Briefly, we found three LAEs in which 
we identified \Ha\ emission. In this paper, we make use of the spectra of 
the three LAEs; CDFS-3865, CDFS-6482, and SSA22-8043 
(Table \ref{tbl:sum_data}). Figure \ref{fig:CM} shows the distributions of 
the LAEs presented in this paper on the $U-$NB387 versus NB387 color 
magnitude diagram.

\subsection{NIRSPEC Observation} \label{ssec:data_nirspec_obs}

The observations were carried out on 9--10 February 2011. Both nights were 
photometric. We observed in low-resolution mode, with a slit width of 
$0\farcs 76$ and a slit length of $42\arcsec$%
\footnote{
The effective slit length is $\sim 38\arcsec$, since we did not 
place objects in $<2\arcsec$ from the edges of the slit.
}. We observed all six candidate with Nirspec-6 filter (hereafter referred 
to as $K$ band). In addition, we observed COSMOS-30679 with Nirspec-5 
($H$ band) and Nirspec-3 ($J$ band), and CDFS-3865 with $J$ band. The 
wavelength ranges of $K$, $H$, and $J$ bands are $1.88$--$2.31\,\mu$m, 
$1.47$--$1.76\,\mu$m, and $1.15$--$1.36\,\mu$m, respectively for our 
configurations. The resolution in the $K$ band is $R\sim 1500$. Exposure 
times for each object are given in Table \ref{tbl:sum_data}.

\begin{deluxetable*}{lcccccc}
\tablecolumns{7}
\tabletypesize{\scriptsize}
\tablecaption{Fluxes of nebular lines determined from NIR Spectroscopy%
\label{tbl:flux_lines}}
\tablewidth{400pt}
\tablehead{%
\colhead{Object} &
\colhead{\OII$\lambda3727$} &
\colhead{\Hb} &
\colhead{\OIII$\lambda4959$} &
\colhead{\OIII$\lambda5007$} &
\colhead{\Ha} &
\colhead{\NII$\lambda6484$}  
}
\startdata
COSMOS-08501  & 
 \nodata & 
 \nodata & 
 \nodata & 
 \nodata & 
 $1.91\pm 0.36$ & 
 $<0.36$ \\
COSMOS-13636  & 
 \nodata &
 \nodata &
 \nodata &
 \nodata &
 $2.71\pm 0.38$ &
 $<0.38$ \\
COSMOS-30679  & 
 $1.04\pm 0.26$ &
 \nodata\tablenotemark{(\dag)} &
 \nodata\tablenotemark{(\dag)} &
 $4.11\pm 0.41$ &
 $3.11\pm 0.27$ &
 $<0.27$ \\
COSMOS-43982  & 
 \nodata &
 \nodata &
 \nodata &
 \nodata &
 $6.63\pm 0.58$ &
 $4.14\pm 0.58$ \\
{\sl Composite}\tablenotemark{($\star$)} & 
 \nodata &
 \nodata &
 \nodata &
 \nodata &
 $2.64\pm 0.20$ &
 $<0.20$ \\
\\
CDFS-3865  & 
 $2.53\pm 0.41$ &
 $13.0\pm 4.1$ &
 $19.0\pm 3.5$ &
 $53.6\pm 3.5$ &
 $38.5\pm 2.4$ &
 $<2.4$ \\
CDFS-6482  & 
 \nodata &
 \nodata &
 \nodata &
 $23.8\pm 2.7$ &
 $9.38\pm 1.71$ &
 $<1.71$ \\
SSA22-8043  & 
 \nodata &
 \nodata &
 \nodata &
 $11.0\pm 4.2$ &
 $14.7\pm 4.3$ &
 $<4.3$ 
\enddata
\tablecomments{
Fluxes and their $1\sigma$ errors are given in unit of $10^{-17}$\,\ergscm.
For lines with less than the $3\sigma$ detection level, we list their 
$1\sigma$ upper-limits.
($\star$) Composite spectrum of the four objects (\S\ref{ssec:AGN}). 
(\dag) These lines suffer badly from OH-lines subtraction errors.
}
\end{deluxetable*}

Our science targets were acquired using the invisible object acquisition 
procedures. In this mode, an alignment star brighter than 
$K_{\rm Vega}\sim 18$ was placed in the slit simultaneously with a science 
target. We first acquired the star at the center of the slit, then we 
nodded the telescope so that both the science target and the star were 
on the slit with the same distance from the slit center. For CDFS-3865 
and HHDFN-18431, since we could not find any stars to place on the slit 
along with them, we first acquired the nearest star to the target (at 
distances of $68\arcsec$ for CDFS-3865 and $39\arcsec$ for HHDFN-18431), 
then nodded the telescope with an offset, which was calculated from the 
WCS difference between the star and the science target. For HHDFN-18325, 
although a star was found at a distance of $32\arcsec$, which is smaller 
than the slit length, half of the observation was done without the star 
in order to avoid the persistence caused by the previous target alignment 
star. During an exposure, we manually guided the slit using the 
slit-viewing camera; if the star started to drift out of the slit, we 
manually moved the slit so that the star (and thus the invisible science 
target) stayed in the slit throughout each exposure.

Standard stars, which were selected from the Hipparcos catalog to have A0V 
spectral type and similar airmass to the science targets, were observed at 
the beginnings and the ends of the nights. The calibration data were taken 
in an ABBA position pattern, while the science data in AB position pattern.

\subsection{NIRSPEC Data Reduction} \label{ssec:data_reduction}

\begin{figure*}
\epsscale{0.9}
\plotone{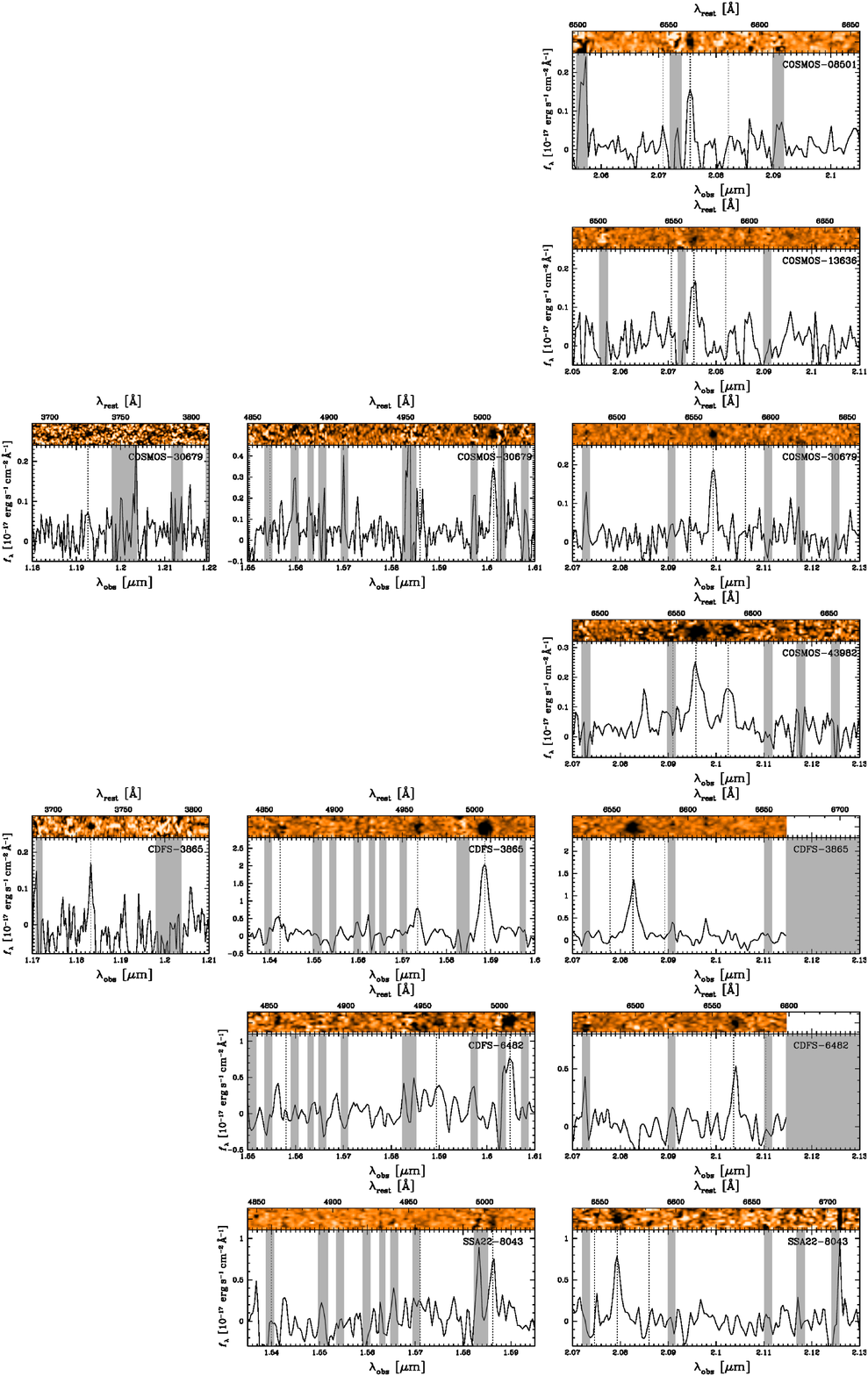}
\caption{
From left to right, $J$, $H$, and $K$ band spectra of the LAEs. In each 
panel, the 2D and 1D spectra are shown in the top and bottom, respectively. 
The vertical dotted lines in the 1D spectra show the expected locations of 
nebular emission lines; \OII$\lambda3727$, \Hb, 
\OIII$\lambda\lambda4959,5007$, \NII$\lambda6548$, \Ha, and 
\NII$\lambda6584$ based on the peak of the detected \Ha\ emission lines.
The gray shaded areas overlaid on each 1D spectrum highlight spectral 
regions strongly affected by OH-airglow. For CDFS objects, their $K$ band 
spectra at $2.115\,\mu$m and longer are not obtained due to a lack of 
sensitivity of MMIRS, and also shaded with gray. 
\label{fig:spectra}}
\end{figure*}

We used the Keck IRAF-based WMKONSPEC package%
\footnote{
\url{http://www2.keck.hawaii.edu/inst/nirspec/wmkonspec.html}
}
to reduce the data. We also used IRAF scripts that were originally written 
for reducing near-infrared multi-object spectroscopic data from 
Subaru/MOIRCS (MCSMDP; \citealt{yoshikawa2010}).

The data reduction process included bias subtraction, flat fielding, 
distortion correction, wavelength calibration, cosmic ray rejection, (A-B) 
sky subtraction, residual background subtraction, image shifting, and 
stacking. We used the bias and flat data that were obtained in the same 
night as the science data. We performed wavelength calibration using 
OH-lines by comparing them with an OH-line list \citep{rousselot2000}.
For cosmic ray rejection, we used {\it LA.COSMIC} \citep{vandokkum2001}.
The sky and OH-lines of an A-position image were roughly removed by 
subtracting an average B-position image created from the previous and 
following images. We then removed the residual sky by subtracting a 
$9$-th order polynomial fit in the spatial direction after masking columns 
of positive and negative parts caused by the alignment star and the object. 
Individual images were shifted in spatial direction so that the object is 
in the same position both in A-position and B-position images. The offset 
values were derived from the position differences of the alignment stars. 
When no star was observed simultaneously, we defined the offset values as 
the nod separation size. Finally, we stacked the position-matched 
individual images to create the two-dimensional (2D) spectra.

We obtained flux solutions by comparing spectra of the A0V standard stars 
and a model, which was created by a stellar spectral synthesis program 
(SPECTRUM; \citealt{gray1994}) based on the \citet{kurucz1993}'s atmosphere 
models. The model spectra were then normalized so that their $J$, $H$, 
and $K$ band magnitudes matched with those of standard stars, whose 
photometry was obtained from the Two Micron All Sky Survey (2MASS) All-Sky 
Point Source Catalog.
The one-dimensional (1D) spectra were extracted by summing up $6$--$10$ 
pixels in the spatial direction. The lengths of columns were determined 
based on the seeing conditions; about two times the seeing size was used. 
We also confirmed that the sizes were large enough to detect most of the 
signal, and maximize the signal-to-noise (S/N) ratio. The 1D and 2D 
spectra are shown in Figure \ref{fig:spectra}. 
We note in addition that the science targets and the standard stars used 
for the flux calibrations were observed and reduced with almost the same 
conditions, thus effects of slit losses were also corrected in the 
procedures, since the standard spectra extracted in the procedure were 
normalized to have the total magnitude of that star. LAEs typically have 
half-light radii $\lesssim 1.5$\,kpc ($\lesssim 0\farcs 18$ at $z=2.2$; 
\citealt{bond2009}), which is much smaller than the seeing size. Therefore, 
we cannot resolve LAEs with our observations, and assuming the PSF profile 
for the LAEs is reasonable.

Emission line fluxes were measured by fitting a Gaussian profile to each 
line with the IRAF task \verb+splot+. The sky noise level was estimated 
in the following manner; we prepared an aperture box, which had 
approximately twice the seeing size in spatial direction and twice the 
FWHM of the best fit Gaussian to the emission line in wavelength direction. 
We spread more than $100$ aperture boxes around the emission line after 
masking pixels heavily contaminated by OH-lines, and measured their 
photon counts. We then fit the histogram of the counts with a Gaussian, 
and regarded its $\sigma$ as the $1\sigma$ fluctuation for the aperture 
used to measure the emission line.
The line flues and their $1\sigma$ errors are summarized in Table 
\ref{tbl:flux_lines}.

\begin{figure}
\epsscale{1.15}
\plotone{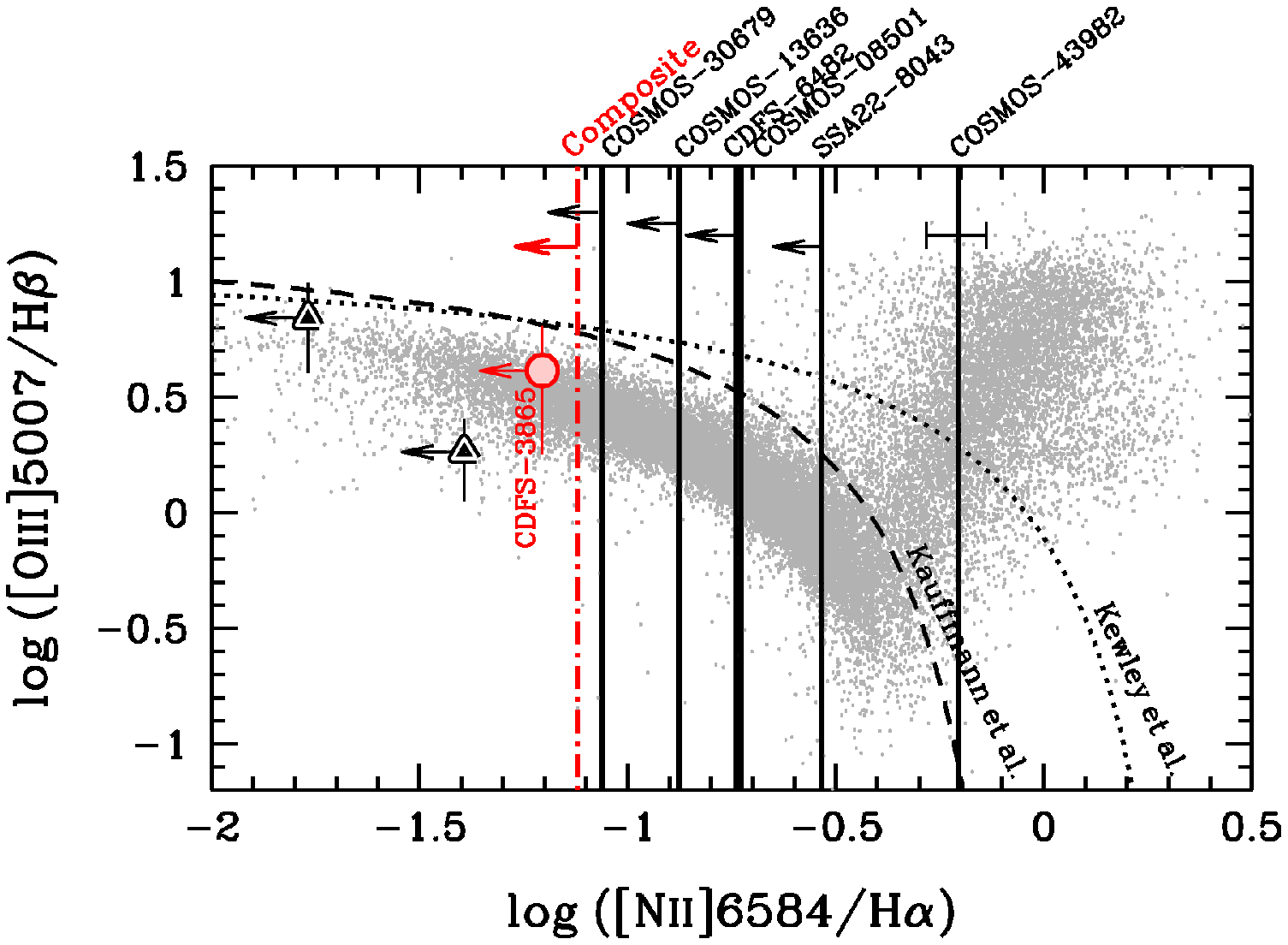}
\caption{
BPT-diagram \citep{BPT1981}. 
The dotted and dashed curves show the dividing line between star-forming 
galaxies and AGNs defined by \citet{kewley2001} and \citet{kauffmann2003}, 
respectively. CDFS-3865 is shown with the red circle, while the other 
LAEs are shown with the vertical solid lines, due to the lacks of \OIII\ 
and/or \Hb. Since we do not identify \NII\ except for COSMOS-43982, 
upper-limits of \NII/Ha\ are given. The \NII/Ha\ ratio for COSMOS-43982 
is given with a horizontal error bar. The red dot-dashed vertical line 
indicates the upper-limit of \NII/\Ha\ for the composite spectrum of 
NIRSPEC-detected LAEs. The two triangles represent LAEs at $z\sim 2.3$ 
and $2.5$ from \citet{finkelstein2011}. The gray small points are randomly 
selected objects from the SDSS spectroscopic sample. 
\label{fig:BPT}} 
\end{figure}

\section{Emission Line Detections} \label{sec:line_measure}

\subsection{\Ha\ detection} \label{ssec:line_ha}

We detect significant \Ha\ emission in $K$ band 2D spectra for COSMOS-08501, 
COSMOS-13636, COSMOS-30679, COSMOS-43982, CDFS-3865, CDFS-6482, and 
SSA22-8043, but we do not identify any emission lines for the other two 
LAEs, HHDFN-18325 and HHDFN-18431. Both LAEs are expected to have strong 
\Lya, hence strong \Ha. A possible reason why we do not detect lines for 
HHDFN-18431 is that no alignment star was observed simultaneously, and 
the object may have drifted significantly out of the slit. The same issue 
may explain the non-detection of \Ha\ for HHDFN-18325, since half of the 
observation was done without an alignment star 
(\S\ref{ssec:data_nirspec_obs}). Since we have not yet obtained \Lya\ 
spectra for the two objects, the non-detections may be alternatively due 
to the mis-selection of LAEs. Future optical spectroscopy is needed to 
resolve this issue. In the following sections, we use the seven LAEs with 
reliable \Ha\ detection.

\subsection{Other emission line detections} \label{ssec:line_other}

\begin{figure}
\epsscale{0.9}
\plotone{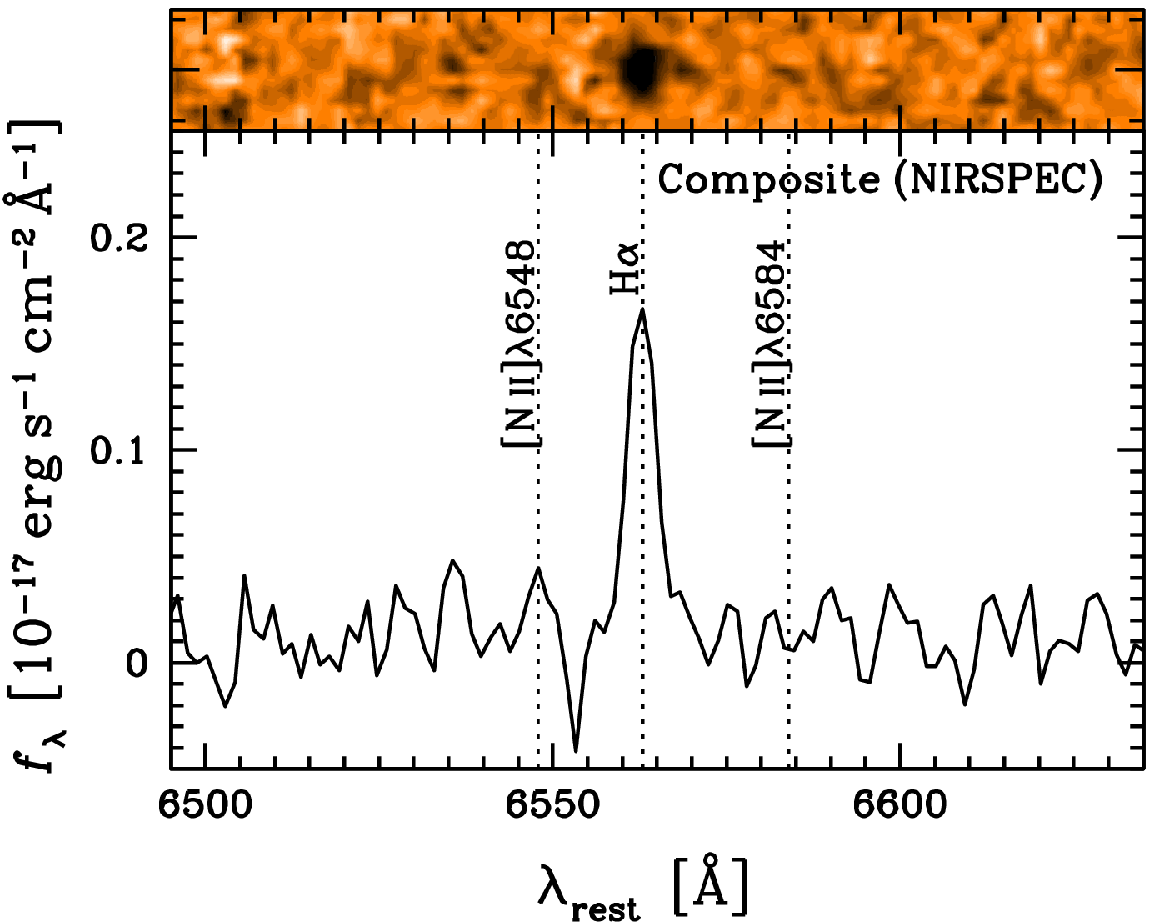}
\caption{
$K$ band composite spectrum of the NIRSPEC-detected objects. 
\label{fig:spec_ALL_n6}}
\end{figure}

\begin{deluxetable*}{lcccccccccccc}
\tablecolumns{13}
\tabletypesize{\scriptsize}
\tablecaption{Broadband Photometry of the LAEs
\label{tbl:BB_photometry}}
\tablehead{
\colhead{COSMOS sample\tablenotemark{(1)}} &
\colhead{$B$} &
\colhead{$V$} &
\colhead{$r'$} &
\colhead{$i'$} &
\colhead{$z'$} &
\colhead{$J$} &
\colhead{\nodata} &
\colhead{$K_s$} &
\colhead{$[3.6]$} &
\colhead{$[4.5]$} &
\colhead{$[5.8]$} &
\colhead{$[8.0]$} 
}
\startdata
COSMOS-08501 &
$25.86$ &
$25.91$ &
$26.08$ &
$25.88$ &
$25.81$ &
$98.45$ &
\nodata &
$25.64$ &
$99.99$ &
$99.99$ &
$99.99$ &
$99.99$
\\
COSMOS-13636 &
$24.43$ &
$24.21$ &
$24.35$ &
$24.19$ &
$24.24$ &
$23.10$ &
\nodata &
$23.43$ &
$24.10$ &
$23.75$ &
$99.99$ &
$99.99$
\\
COSMOS-30679 &
$24.05$ &
$23.12$ &
$22.91$ &
$22.46$ &
$22.33$ &
$21.15$ &
\nodata &
$21.82$ &
$22.12$ &
$22.57$ &
$99.99$ &
$23.06$
\\
COSMOS-30679\tablenotemark{(\dag)} &
$24.76$ &
$23.82$ &
$24.44$ &
$24.09$ &
$23.49$ &
$22.31$ &
\nodata &
$23.29$ &
\nodata &
\nodata &
\nodata &
\nodata
\\
COSMOS-43982 &
$25.00$ &
$24.38$ &
$24.48$ &
$23.99$ &
$23.73$ &
$21.89$ &
\nodata &
$21.62$ &
$21.20$ &
$21.02$ &
$20.69$ &
$20.75$
\\
(limitmag) &
$(29.13)$ &
$(28.18)$ &
$(28.33)$ &
$(27.87)$ &
$(26.89)$ &
$(24.17)$ &
\nodata &
$(24.84)$ &
$(25.05)$ &
$(24.25)$ &
$(21.90)$ &
$(20.63)$ 
\\
(limitmag)\tablenotemark{(\dag)} &
$(28.76)$ &
$(26.24)$ &
$(25.79)$ &
$(25.34)$ &
$(24.85)$ &
$(23.68)$ &
\nodata &
$(24.63)$ &
\nodata &
\nodata &
\nodata &
\nodata 
\\
\\
\colhead{CDFS sample\tablenotemark{(2)}} &
\colhead{$B$} &
\colhead{$V$} &
\colhead{$R$} &
\colhead{$I$} &
\colhead{$z'$} &
\colhead{$J$} &
\colhead{$H$} &
\colhead{$K$} &
\colhead{$[3.6]$} &
\colhead{$[4.5]$} &
\colhead{$[5.8]$} &
\colhead{$[8.0]$} 
\\
\hline
CDFS-3865 & 
$23.01$ &
$22.94$ &
$22.92$ &
$23.14$ &
$22.93$ &
$22.73$ &
$22.27$ &
$22.38$ &
$22.82$ &
$22.82$ &
$22.51$ &
$23.00$
\\
CDFS-6482 &
$23.93$ &
$23.87$ &
$23.78$ &
$23.95$ &
$23.67$ &
$23.50$ &
$23.36$ &
$23.07$ &
$22.88$ &
$22.83$ &
$22.34$ &
$99.99$
\\
(limitmag) &
$(28.32)$ &
$(27.85)$ &
$(27.82)$ &
$(26.14)$ &
$(25.64)$ &
$(24.57)$ &
$(24.57)$ &
$(23.97)$ &
$(26.23)$ &
$(25.68)$ &
$(23.66)$ &
$(23.43)$
\\
\\
\colhead{SSA22 sample\tablenotemark{(3)}} &
\colhead{$B$} &
\colhead{$V$} &
\colhead{$R$} &
\colhead{$i'$} &
\colhead{$z'$} &
\colhead{$J$} &
\colhead{\nodata} &
\colhead{$K$} &
\colhead{$[3.6]$} &
\colhead{$[4.5]$} &
\colhead{$[5.8]$} &
\colhead{$[8.0]$} 
\\
\hline
SSA22-8043 & 
$24.80$ &
$24.63$ &
$25.54$ &
$24.64$ &
$24.59$ &
$24.10$ &
\nodata &
$21.79$ &
$23.33$ &
$23.14$ &
$26.58$ &
$21.15$ 
\\
(limitmag) &
$(27.85)$ &
$(27.99)$ &
$(28.03)$ &
$(27.75)$ &
$(27.16)$ &
$(24.74)$ &
\nodata &
$(22.78)$ &
$(25.40)$ &
$(23.64)$ &
$(22.71)$ &
$(20.95)$
\enddata
\tablenotetext{(1)}{%
$BVr'i'z'$ data are obtained from Subaru/Suprime-Cam, 
$J$ data from UKIRT/WFCAM, 
$K_s$ data from CFHT/WIRCAM, and 
$[3.6]-[8.0]$ data from Spitzer/IRAC.
All data are collected from the COSMOS Archive.
}
\tablenotetext{(2)}{%
$BVRI$ data are obtained from MPG/ESO 2.2m/WFI, 
$z'$ data from CTIO/MOSAIC II, 
$JK$ data from CTIO/ISPI, 
$H$ data from NTT/SOFI, and 
$[3.6]-[8.0]$ data from  Spitzer/IRAC.
Optical and NIR data are collected from 
the MUSYC Public Data Release \citep{gawiser2006}, 
and Spitzer data from the SIMPLE Legacy II project.
}
\tablenotetext{(3)}{%
$BVRi'z'$ data are obtained from Subaru/Suprime-Cam, 
$J$ data from KPNO/NEWFIRM, 
$K$ data from UKIRT/WFCAM, and 
$[3.6]-[8.0]$ data from Spitzer/IRAC.
Optical data are provided by T. Hayashino 
(see also \citealt{hayashino2004}), 
$J$ data are obtained from the New\Ha\ survey \citep{ly2011,lee2012}.
$K$ data are collected from the WFCAM data access page, 
and Spitzer data from the Spitzer Heritage Archive.
}
\tablenotetext{($\dagger$)}{%
Values obtained after removing the contribution from an 
adjacent object (\S\ref{ssec:SEDfit}).
The limiting magnitudes are calculated by adding the photometric errors 
and additional errors (residuals after subtracting GALFIT models) in 
quadrature. The limiting magnitudes are thus shallower than those given 
above (photometric only).
}
\tablecomments{
Broadband photometry of the LAEs.
All magnitudes are total magnitudes.
99.99 mag means no signal detected.
Magnitudes in parentheses are $1\sigma$ uncertainties 
adopted in SED fitting. 
}
\end{deluxetable*}

For CDFS-3865, we additionally obtain \OII$\lambda3727$, \Hb, and 
\OIII$\lambda\lambda 4959, 5007$ emission lines. Combined with the 
\Ha-detection, a full suite of prominent rest-frame optical nebular lines 
are thus obtained for this object. We note that the \OII\ flux may be 
underestimated due to the possible flux loss caused by the guiding drift 
issue known to NIRSPEC. Since CDFS-3865 was observed with NIRSPEC without 
any alignment star simultaneously, the maximum flux loss caused by the 
guiding drift ($\sim 1\farcs 5$ per hour; private communication with 
NIRSPEC support scientists) could be comparable to the observed value. 
However, we do not know the drift direction, and the true flux loss cannot 
be estimated. Therefore, we do not correct for any flux loss for \OII.
We emphasize that the possible factor$\lesssim 2$ uncertainties in \OII\ 
flux do not change our main results (\S\ref{ssec:qion_Z}). For COSMOS-30679, 
we obtain \OII\ and \OIII$\lambda 5007$ emission lines, while we do not 
identify \Hb\ and \OIII$\lambda 4959$, because they fall in a dense OH-line 
wavelength range. We thus cannot place meaningful upper limits on their 
fluxes. When fluxes for these lines are required for inferring physical 
properties, we assume a value of $0.28$ for the \OIII$\lambda 4959/5007$ 
ratio (e.g., \citealt{richard2011})
and calculate the \Hb\ flux based on the observed \Ha\ flux, the 
intrinsic \Ha/\Hb\ ratio assuming Case B recombination (2.86; 
\citealt{osterbrock1989}), and the dust extinction estimated from SED 
fitting. We have checked for CDFS-3865 that the \Hb\ flux estimated in this 
way is consistent with the really observed \Hb\ flux (see also 
\S\ref{ssec:SEDfit}).
We emphasize that the two LAEs, CDFS-3865 and COSMOS-30679, are the first 
\OII-detected LAEs individually at high-$z$. The \OII+\OIII\ lines provide 
the simultaneous determinations of ionization state and oxygen abundance 
(\S\ref{ssec:qion_Z}).

For CDFS-6482 and SSA22-8043, we detect \OIII$\lambda 5007$ while we do not 
identify \Hb\ and \OIII$\lambda 4959$, probably due to the contamination by 
OH-lines and the limited sensitivity of the instrument (More detailed 
descriptions about the MMIRS objects are provided by 
\citealt{hashimoto2013}). 
For COSMOS-43982, we clearly detect \NII$\lambda6584$ emission line, which 
is not identified for the other LAEs. The \NII-detection suggests a 
non-negligible contribution of AGN to the emission lines for the object. 
Further investigations of AGN contamination are discussed in 
\S\ref{ssec:AGN}.

\section{Physical Properties of LAEs} \label{sec:properties}

\subsection{Removal of objects with AGN} 
\label{ssec:AGN}

AGN, as well as star-formation, can produce large amounts of ionizing 
photons, and show strong \Lya. Since our interest is star-forming LAEs, 
we need to remove LAEs with AGN activity from the sample.

During the LAE sample selection, bright LAEs with AGN signature have been 
removed (see \citealt{nakajima2012}). Obvious AGNs are therefore excluded
from the sample. However, the procedure is not always perfect, and can miss 
objects with relatively weak AGN activity. In order to assess possible 
contamination of AGN in our LAE sample, we plot our LAEs on a BPT diagram 
(Figure \ref{fig:BPT}; \citealt{BPT1981}) which is widely used to separate 
star forming galaxies from AGN. In Figure \ref{fig:BPT}, the underlying 
small gray points denote the spectroscopic objects from the Sloan Digital 
Sky Survey (SDSS; \citealt{york2000})%
\footnote{
We use a part of the spectroscopic data taken from the MPA-JHU DR7 release 
of spectrum measurements: \url{http://www.mpa-garching.mpg.de/SDSS/DR7/}
}. 
The two curves shown in Figure \ref{fig:BPT} are the empirical demarcations 
\citep{kewley2001,kauffmann2003}. Since only one LAE, CDFS-3865, has both 
\OIII\ and \Hb, all the other LAEs are shown with vertical lines. Due to 
the non-detection of \NII, the vertical lines show upper-limits of \NII/\Ha\ 
except for COSMOS-43982, which has an individual \NII\ detection. Based on 
its relatively high \NII/\Ha\ ratio, COSMOS-43982 may be contaminated by an 
AGN, though there remains a possibility that this LAE may be a star-formation 
dominated galaxy with a very small \OIII/\Hb\ ratio. For this analysis, we 
choose to be conservative and regard the object as a candidate possessing 
AGN activity. The properties of COSMOS-43982 are considered and interpreted 
carefully in the following sections.

\begin{figure*}
\epsscale{1.10}
\plotone{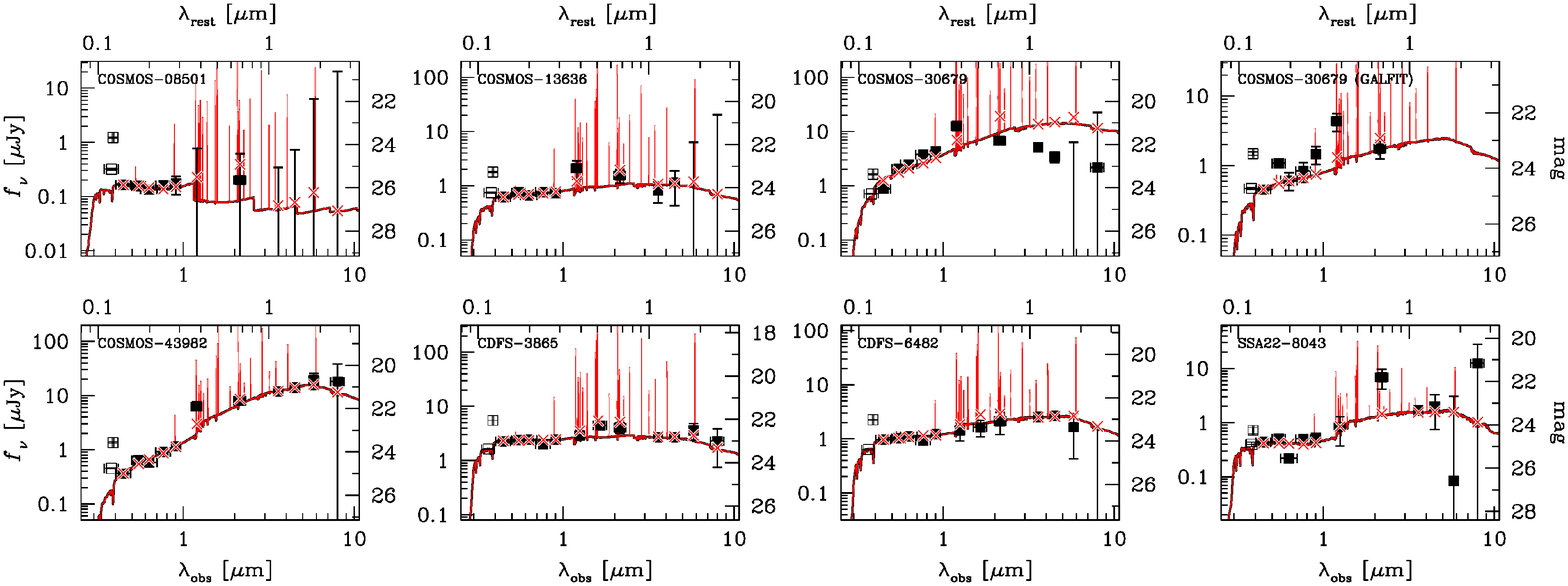}
\caption{
Results of SED fitting. 
The filled squares show the observed flux densities used for the fitting 
($B$, $V$, $r$, $i$, $z$, $J$, ($H$), $K$, $[3.6]$, $[4.5]$, $[5.8]$, and 
$[8.0]$), while the open squares indicate those we omit for the fitting 
($U$ and NB387), due to the unknown flux contributions of IGM absorption. 
The red lines show the best-fit model spectra, and the 
red crosses correspond to the best-fit flux densities.
\label{fig:SEDfit}}
\end{figure*}

\begin{deluxetable*}{lccccr}
\tablecolumns{6}
\tabletypesize{\scriptsize}
\tablecaption{Physical Properties from SED Fitting%
\label{tbl:SEDfit}}
\tablewidth{350pt}
\tablehead{%
\colhead{Object} &
\colhead{$\log(M_{\star})$} &
\colhead{$E(B-V)$} &
\colhead{$\log({\rm age})$} &
\colhead{$\log({\rm SFR})$} &
\colhead{$\chi^2$} \\
\colhead{} &
\colhead{($M_{\odot}$)} & 
\colhead{(mag)} &
\colhead{(yr)} &
\colhead{($M_{\odot}/{\rm yr}$)} &
\colhead{} 
}
\startdata
COSMOS-08501 &
 $7.84^{+1.21}_{-0.27}$ &
 $0.08^{+0.04}_{-0.08}$ &
 $6.16^{+2.75}_{-1.06}$ &
 $1.68^{+1.24}_{-1.44}$ &
 $1.825$ \\
COSMOS-13636 &
 $9.12^{+0.13}_{-0.14}$ &
 $0.18^{+0.01}_{-0.01}$ &
 $7.34^{+0.23}_{-0.22}$ &
 $1.80^{+0.08}_{-0.09}$ &
 $22.356$ \\
COSMOS-30679 &
 $10.34^{+0.00}_{-0.00}$ &
 $0.40^{+0.00}_{-0.00}$ &
 $7.26^{+0.00}_{-0.00}$ &
 $3.10^{+0.00}_{-0.00}$ &
 $7062.297$ \\
COSMOS-30679\tablenotemark{(\dag)} &
 $9.74^{+0.26}_{-0.52}$ &
 $0.24^{+0.04}_{-0.04}$ &
 $8.01^{+0.45}_{-0.79}$ &
 $1.79^{+0.29}_{-0.19}$ &
 $32.068$ \\
COSMOS-43982 &
 $10.80^{+0.01}_{-0.06}$ &
 $0.40^{+0.02}_{-0.01}$ &
 $8.51^{+0.05}_{-0.15}$ &
 $2.37^{+0.08}_{-0.04}$ &
 $71.642$ \\
CDFS-3865 &
 $9.50^{+0.01}_{-0.03}$ &
 $0.14^{+0.00}_{-0.00}$ &
 $7.36^{+0.02}_{-0.04}$ &
 $2.17^{+0.01}_{-0.01}$ &
 $32.170$ \\
CDFS-6482 &
 $9.80^{+0.06}_{-0.05}$ &
 $0.15^{+0.02}_{-0.02}$ &
 $8.16^{+0.15}_{-0.15}$ &
 $1.71^{+0.09}_{-0.09}$ &
 $12.374$ \\
SSA22-8043 &
 $10.07^{+0.06}_{-0.06}$ &
 $0.03^{+0.02}_{-0.01}$ &
 $9.44^{+0.00}_{-0.10}$ &
 $0.76^{+0.07}_{-0.03}$ &
 $112.288$ 
\enddata
\tablecomments{
Physical properties and their $1\sigma$ errors of LAEs from SED fitting.
Stellar metallicity is fixed to $0.2\,Z_{\odot}$. Escape fraction of 
ionizing photons is fixed to $0.05$.
(\dag) Values obtained after removing the contribution from an adjacent 
object (\S\ref{ssec:SEDfit}).
}
\end{deluxetable*}

For CDFS-3865, thanks to its relatively strong constraint on \NII/\Ha, we can 
see that the object is along the star forming sequence and has negligible AGN 
activity, similar to two other LAEs with spectroscopic follow-up at similar 
redshifts \citep{finkelstein2011}. Although the other LAEs have no constraint 
on their $y$-axis values, AGN contributions to them are assumed to be minimal, 
since there are few AGNs in the range $\log$(\NII/\Ha) $\lesssim -0.5$, as 
suggested from the underlying SDSS galaxies. In order to check the assumption, 
we stack the $K$ band spectra of the four NIRSPEC-detected LAEs%
\footnote{
We do not use the MMIRS spectra due to their worse sensitivity and spectral 
resolution.
} with inverse-variance weights. The composite spectrum is shown in Figure 
\ref{fig:spec_ALL_n6}, and provides an average, deeper constraint on 
\NII/\Ha\ ratio for the LAEs. The red dot-dashed line in Figure \ref{fig:BPT}
shows the upper-limit of \NII/\Ha\ ratio constrained by the composite 
spectrum. The line is further away from the area occupied with AGNs. 
Therefore, AGN activity is on average negligible for the LAEs. In the 
following sections, we regard all the LAEs except for COSMOS-43982 as 
galaxies dominated by star-formation.

\subsection{Stellar Mass and E(B-V) from SED fitting} 
\label{ssec:SEDfit}

\begin{figure}
\epsscale{1.15}
\plotone{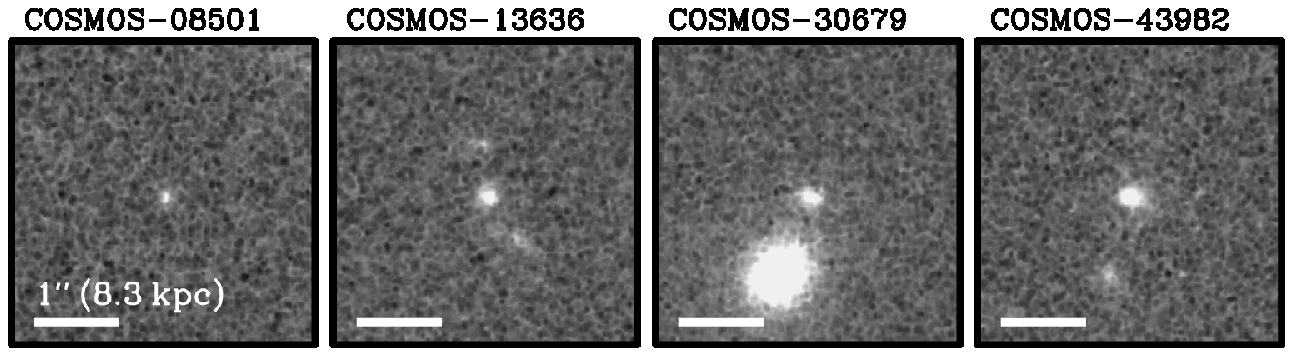}
\caption{
HST/ACS F814W images for COSMOS objects. White tick indicates $1$\,arcsec, 
corresponding to $\sim 8.3$\,kpc at $z=2.2$. North is up and east is to the 
left.
\label{fig:acs_I}}
\end{figure}

We perform SED fitting to broadband photometry to infer properties of the 
stellar populations. The majority of images are collected from publicly 
available databases (see notes in Table \ref{tbl:BB_photometry}). The 
optical and NIR photometry is done with a $2\farcs 0$ diameter aperture 
using the double-image mode of SExtractor \citep{BA1996}. The aperture 
magnitudes are then converted into total magnitudes using aperture 
correction values, which are estimated from differences between aperture 
magnitudes and \verb+MAG_AUTO+ values for point sources. For the IRAC 
imaging, we use a $3\farcs 0$ diameter aperture and \citet{yan2005}'s 
aperture correction values. Table \ref{tbl:BB_photometry} summarizes the 
results.

\begin{deluxetable*}{p{67pt}cccccccccc}
\tablecolumns{11}
\tabletypesize{\scriptsize}
\tablecaption{Physical Properties from the Nebular lines%
\label{tbl:properties_spec}}
\tablehead{%
\colhead{Object} &
\multicolumn{2}{c}{\Oabundance} &
\colhead{$q_{ion}$} &
\colhead{SFR} &
\colhead{SFR$_{\rm cor}$} &
\colhead{EW(\Ha)} &
\colhead{$L$(\Ha)} &
\colhead{EW(\Lya)} &
\colhead{$L$(\Lya)} &
\colhead{AGN} \\
\colhead{} &
\colhead{$N2$} & 
\colhead{$R23$} & 
\colhead{\tiny ($10^7$\,cm\,s$^{-1}$)} &
\colhead{\tiny ($M_{\odot}\,{\rm yr}^{-1}$)} &
\colhead{\tiny ($M_{\odot}\,{\rm yr}^{-1}$)} &
\colhead{\tiny (\AA)} & 
\colhead{\tiny ($10^{42}$\,\ergs)} & 
\colhead{\tiny (\AA)} & 
\colhead{\tiny ($10^{42}$\,\ergs)} & 
\colhead{\tiny (10)} \\
\colhead{} &
\colhead{\tiny (1)} & 
\colhead{\tiny (2)} & 
\colhead{\tiny (3)} &
\colhead{\tiny (4)} &
\colhead{\tiny (5)} &
\colhead{\tiny (6)} & 
\colhead{\tiny (7)} & 
\colhead{\tiny (8)} & 
\colhead{\tiny (9)} & 
\colhead{}    
}
\startdata
{\scriptsize COSMOS-08501} &
 $<8.73$ &
 \nodata &
 \nodata &
 $5.3^{+1.0}_{-1.0}$ &
 $6.7^{+1.6}_{-2.1}$ &
 $>280$ &
 $0.67\pm 0.13$ &
 $255\pm 26$ & 
 $8.41\pm 0.40$ & 
 $0$ 
 \\
{\scriptsize COSMOS-13636} &
 $<8.61$ &
 \nodata &
 \nodata &
 $7.5^{+1.1}_{-1.1}$ &
 $13.2^{+1.9}_{-1.9}$ &
 $93^{+26}_{-32}$ &
 $0.95\pm 0.13$ &
 $73\pm 5$ & 
 $11.35\pm 0.50$ & 
 $0$ 
 \\
{\scriptsize COSMOS-30679}\tablenotemark{(\dag)}  &
 $<8.46$ &
 $8.18^{+0.28}_{-0.28}$ &
 $8^{+10}_{-4}$ &
 $9.0^{+0.8}_{-0.8}$ &
 $18.6^{+2.6}_{-2.6}$ &
 $93^{+25}_{-33}$ &
 $1.14\pm 0.10$ &
 $87\pm 7$ & 
 $8.47\pm 0.65$ & 
 $0$ 
 \\
{\scriptsize COSMOS-43982} &
 \nodata &
 \nodata &
 \nodata &
 $19.0^{+1.7}_{-1.7}$ &
 $65.7^{+6.7}_{-6.0}$ &
 $41^{+4}_{-4}$ &
 $2.41\pm 0.21$ &
 $105\pm 8$ & 
 $11.00\pm 0.47$ & 
 $1$ 
 \\
{\scriptsize CDFS-3865} &
 $<8.35$ &
 $7.84^{+0.24}_{-0.25}$ &
 $25^{+17}_{-8}$ &
 $107.9^{+6.7}_{-6.7}$ &
 $166.0^{+10.3}_{-10.3}$ &
 $813^{+175}_{-216}$ &
 $13.65\pm 0.85$ &
 $64\pm 29$ & 
 $29.79\pm 4.93$ & 
 $0$ 
 \\
{\scriptsize CDFS-6482} &
 $<8.72$ &
 \nodata &
 \nodata &
 $27.2^{+5.0}_{-5.0}$ &
 $43.1^{+8.2}_{-8.2}$ &
 $261^{+103}_{-149}$ &
 $3.45\pm 0.63$ &
 $75\pm 52$ & 
 $15.40\pm 8.09$ & 
 $0$ 
 \\
{\scriptsize SSA22-8043} &
 $<8.90$ &
 \nodata &
 \nodata &
 $41.0^{+12.0}_{-12.0}$ &
 $44.4^{+13.2}_{-13.1}$ &
 $120^{+53}_{-69}$ &
 $5.19\pm 1.52$ &
 $28\pm 4$ & 
 $2.22\pm 0.38$ & 
 $0$ 
\enddata
\tablecomments{
(1) Oxygen abundance derived from $N2$-index (\S\ref{ssec:qion_Z}). 
If \NII\ is undetected, $1\sigma$ upper-limit is given. 
(2) Oxygen abundance derived from $R23$-index (\S\ref{ssec:qion_Z}). 
Ionization parameter is derived at the same time.
(3) Ionization parameter ($10^7$\,cm\,s$^{-1}$) estimated from the \OIII/\OII\ 
ratio and $Z(R23)$ (\S\ref{ssec:qion_Z}). 
(4) Star formation rate ($M_{\odot}\,{\rm yr}^{-1}$) derived from uncorrected 
\Ha\ luminosity (\S\ref{ssec:SFR}).
(5) Star formation rate ($M_{\odot}\,{\rm yr}^{-1}$) derived from \Ha\ 
luminosity after correcting for extinction derived from SED fitting 
(\S\ref{ssec:SFR}).
(6) Rest-frame equivalent width of \Ha\ (\AA; \S\ref{ssec:EW_Ha}).
(7) Observed luminosity of \Ha\ ($10^{42}$\,\ergs). 
(8) Rest-frame equivalent width of \Lya\ (\AA; see Table \S\ref{tbl:sum_data}). 
(9) Observed luminosity of \Lya\ ($10^{42}$\,\ergs). 
(10) Flag of AGN (\S\ref{ssec:AGN}): AGN candidates are flagged with "$1$".
(\dag) Values obtained after removing the contribution from an 
adjacent object (\S\ref{ssec:SEDfit}).
}
\end{deluxetable*}

The procedure of the SED fitting is the same as that of \citet{ono2010a}, 
except for fixed redshifts, which are derived from \Ha. Briefly, we use the 
stellar population synthesis model GALAXEV \citep{BC2003} for stellar SEDs, 
and include nebular emission \citep{SdB2009}. We set the escape fraction of 
ionizing photons (\fesc) as $0.05$. Such low \fesc\ has been suggested for 
$z\sim 3$ LBGs (e.g., \citealt{shapley2006}; see also \citealt{iwata2009}). 
Recently, \citet{nestor2013} suggest that faint LAEs may have a higher 
\fesc\ ($\sim 0.10$--$0.30$) in contrast with bright LBGs. We have checked 
that even if we assume \fesc\ $=0.20$, the derived quantities do not change 
significantly. We adopt a Salpeter initial mass function (IMF; 
\citealt{salpeter1955}) with lower and upper mass cut-offs of $0.1$ and 
$100\,M_{\odot}$. We choose constant star-formation history and the stellar 
metallicity $Z=0.2\,Z_{\odot}$. For dust extinction, we use Calzetti's 
extinction law \citep{calzetti2000} on the assumption of 
$E(B-V)_{\rm gas}=E(B-V)_{\star}$ as proposed by \citet{erb2006b}, although 
this assumption is still debatable for high-$z$ galaxies (e.g., 
\citealt{forsterschreiber2009,ly2012})%
\footnote{
Adopting $E(B-V)_{\rm gas}=E(B-V)_{\star}/0.44$ \citep{calzetti2000} 
decreases ionization parameters (\S\ref{ssec:qion_Z}) by a factor of 
$\lesssim 2$, but the changes are smaller than their $1\sigma$ errors. 
Changes in metallicities are also small compared with the errors. 
}. Intergalactic medium (IGM) attenuation is calculated using the 
prescription given by \citet{madau1995}. 
These assumptions are usually used for SED fitting for high-$z$ LAEs. E.g., 
\citet{guaita2011} test three different star-formation histories when 
performing SED fitting to $z\sim 2.1$ stacked LAEs, and find equally good 
fits to the data. The authors also note that SED fitting can relatively 
well constrain stellar mass and dust extinction among the free parameters.
The derived properties are given in Table \ref{tbl:SEDfit}, and the 
best-fit SEDs are shown in Figure \ref{fig:SEDfit}.

One problem is that COSMOS-30679 was found to be blended with another 
brighter adjacent object via examination of the COSMOS HST/ACS F814W 
imaging data (Figure \ref{fig:acs_I}). In order to deblend this object, we 
use the galaxy profile fitting software GALFIT (v3.0; \citealt{peng2010}). 
We first run GALFIT in the $B$ band and fit both objects simultaneously. 
Then, we perform fittings in other bands by fixing parameters (except for 
brightnesses) to those derived in the $B$ band fitting and derive the 
deblended photometry, which is listed in Table \ref{tbl:SEDfit}. We confirm 
that the $i'$ band brightness estimated from the deblended object is 
consistent with the photometry in the HST/ACS F814W high-resolution image 
within their errors. We add residuals after subtracting the fitted profiles 
into the photometric errors in quadrature, so that those errors are used in 
the SED fitting. Since the adjacent object appears dominant at longer 
wavelengths and deblending gets less reliable, we do not perform the 
deblending for the IRAC images. 
We therefore fit the SED of COSMOS-30679 after deblending without the IRAC 
data.
The best fit SEDs of COSMOS-30679 after and before deblending are shown in 
upper right two panels in Figure \ref{fig:SEDfit}. Although the bumpy shape 
of the SED after the deblending suggests additional contamination from the 
adjacent object, the fitting is acceptable (relatively small $\chi^2$ value) 
compared with that obtained before deblending. In the following sections, 
we use the quantities after deblending.

We obtain from our sample stellar masses ranging from 
$7\times 10^{7}\,M_{\odot}$ to $1\times 10^{10}\,M_{\odot}$ and dust extinction 
$E(B-V)=0.03$ to $0.24$ (except for the AGN LAE). These results are 
consistent with the previously recognized trend that LAEs at $z\sim 2$ are 
diverse, unlike higher redshift counterparts (e.g., \citealt{nilsson2011}). 
We also find that dust extinctions estimated from the \Ha/\Hb\ ratio and 
SED fitting are consistent with each other for CDFS-3865 
\citep{hashimoto2013}. This supports our assumption that the dust extinction 
from SED fitting under the assumption of $E(B-V)_{\rm gas}=E(B-V)_{\star}$ 
can be reasonably applied for the extinction correction of nebular lines.

\subsection{Ionization Parameter and Metallicity} \label{ssec:qion_Z}

The ionization state in an \HII\ region is often characterized by the 
ionization parameter, $q_{ion}$%
\footnote{
We use the subscript ``$ion$'' for the ionization parameter to 
distinguish it from the $q$-parameter which stands for the 
effect of dust on \Lya\ compared with UV-continuum 
(\S\ref{sssec:EW_Lya_Ha}).
}, which is the ratio of the mean ionizing photon flux to the mean 
hydrogen atom density. Larger $q_{ion}$ means the \HII\ region is more highly 
ionized, probably achieved by a harder ionizing spectrum. Consequently, 
emission lines with higher ionization potentials are expected to be observed 
more strongly with increasing $q_{ion}$. In this sense, the ionization 
parameter is well determined using ratios of emission lines of different 
ionization stages of the same element, such as the \OIII/\OII\ ratio (e.g., 
\citealt{KD2002}). However, the \OIII/\OII\ ratio depends not only on 
ionization state but also on gas metallicity (Fig. $1$ of \citealt{KD2002}). 
Since the gas temperature decreases as gas metallicity increases, 
the far infrared fine-structure cooling by doubly ionized species 
(i.e., O$^{2+}$) 
becomes efficient when the gas metallicity is high 
(e.g., \citealt{charlot2001,dopita2000}), 
resulting in a decrease of \OIII\ line in optical.
Conversely, metallicity estimates using $R23$-index \citep{pagel1979}: 
\begin{eqnarray}
R23
= \frac{{\rm [O\,{\scriptstyle II}]}\lambda3727
        +{\rm [O\,{\scriptstyle III}]}\lambda\lambda5007,4959}
       {{\rm H}\beta}
\label{eq:R23}
\end{eqnarray}
are affected by the ionization parameter to some extent (e.g., 
\citealt{KD2002,nagao2006}). Therefore, we combine the \OIII/\OII\ ratio and 
$R23$-index (both corrected for dust extinction estimated from SED fitting) 
to estimate its ionization parameter and metallicity at the same time 
iteratively. We assume in this section \fesc\ $=0$ for simplicity. The 
possible effect of non-zero escape fractions will be discussed in 
\S\ref{sssec:high_qion}.

We describe the procedure for CDFS-3865 as an example. As an initial guess 
of metallicity, we use the empirical $R23$ indicator determined by local 
galaxies \citep{maiolino2008}. We obtain two solutions \Oabundance\ 
$=7.45^{+0.24}_{-0.19}$ and $8.66^{+0.20}_{-0.26}$. We then use another 
alternative indicator, $N2$-index defined as \NII$\lambda6584$/\Ha\ 
\citep{maiolino2008}. The upper-limit of  \NII/\Ha\ provides an upper-limit 
of metallicity \Oabundance\ $=8.35$ ($1\sigma$), which removes the 
high-metallicity solution by $R23$-index at the $2.5\sigma$ level. 
Therefore, the empirical relation yields the metallicity \Oabundance\ 
$=7.45^{+0.24}_{-0.19}$.
As the second step, we use this metallicity and the \OIII/\OII\ ratio to 
estimate its ionization parameter. \citet{KD2002} calculate the relations 
between \OIII/\OII\ ratio and ionization parameter with discrete values of 
metallicity. Among them, our initial guess of metallicity corresponds to 
their lowest metallicity ($\sim 0.09\,Z_{\odot}$; \citealt{allende2001}). By 
using the relation, we find that the high \OIII/\OII\ ratio can only be 
reproduced with ionization parameters as high as several times 
$10^8$\,cm\,s$^{-1}$. 
As the third step, we recalculate its metallicity from the \citet{KD2002}'s 
relation ($R23$-index) for the ionization parameter 
$3\times 10^8$\,cm\,s$^{-1}$, and obtain \Oabundance\ $=7.84^{+0.24}_{-0.25}$%
\footnote{
Using the $1.5\times 10^8$\,cm\,s$^{-1}$ relation does not change the 
metallicity estimate significantly ($\sim 0.02$\,dex lower). 
}. Finally, we recalculate its ionization parameter based on the metallicity 
and Eq. (12) of \citet{KD2002}%
\footnote{
Although the original relations are suspended at  
$q_{ion} = 3\times 10^8$\,cm\,s$^{-1}$, we extend their polynomials
toward higher $q_{ion}$ values.
}, and obtain $q_{ion} = 2.5^{+1.7}_{-0.8}\times 10^8$\,cm\,s$^{-1}$. This is 
consistent with the assumed ionization parameter in the third step
($3\times 10^8$\,cm\,s$^{-1}$). We check that further iterations do not 
change our final solutions. To summarize, CDFS-3865 has the ionization 
parameter $q_{ion} = 2.5^{+1.7}_{-0.8}\times 10^8$\,cm\,s$^{-1}$ and the 
metallicity \Oabundance\ $=7.84^{+0.24}_{-0.25}$.

In the same manner, we estimate COSMOS-30679's ionization parameter and 
metallicity to be $q_{ion} = 8^{+10}_{-4}\times 10^7$\,cm\,s$^{-1}$ and 
\Oabundance\ $=8.18^{+0.28}_{-0.28}$, respectively. Since COSMOS-30679 does 
not have \Hb, its strength is inferred from \Ha\ assuming the Case B and 
the dust extinction. In addition, its broadband SED is fitted after being 
deblended with the adjacent object (\S\ref{ssec:SEDfit}). Therefore, the 
object may possess larger uncertainties.

For the LAEs without \OII, we estimate their metallicities by using the 
empirical $N2$-index relation, and no constraint on ionization parameters 
is provided for them.
The metallicities and the ionization parameters are summarized in Table 
\ref{tbl:properties_spec}.

\begin{figure*}
\epsscale{0.85}
\plotone{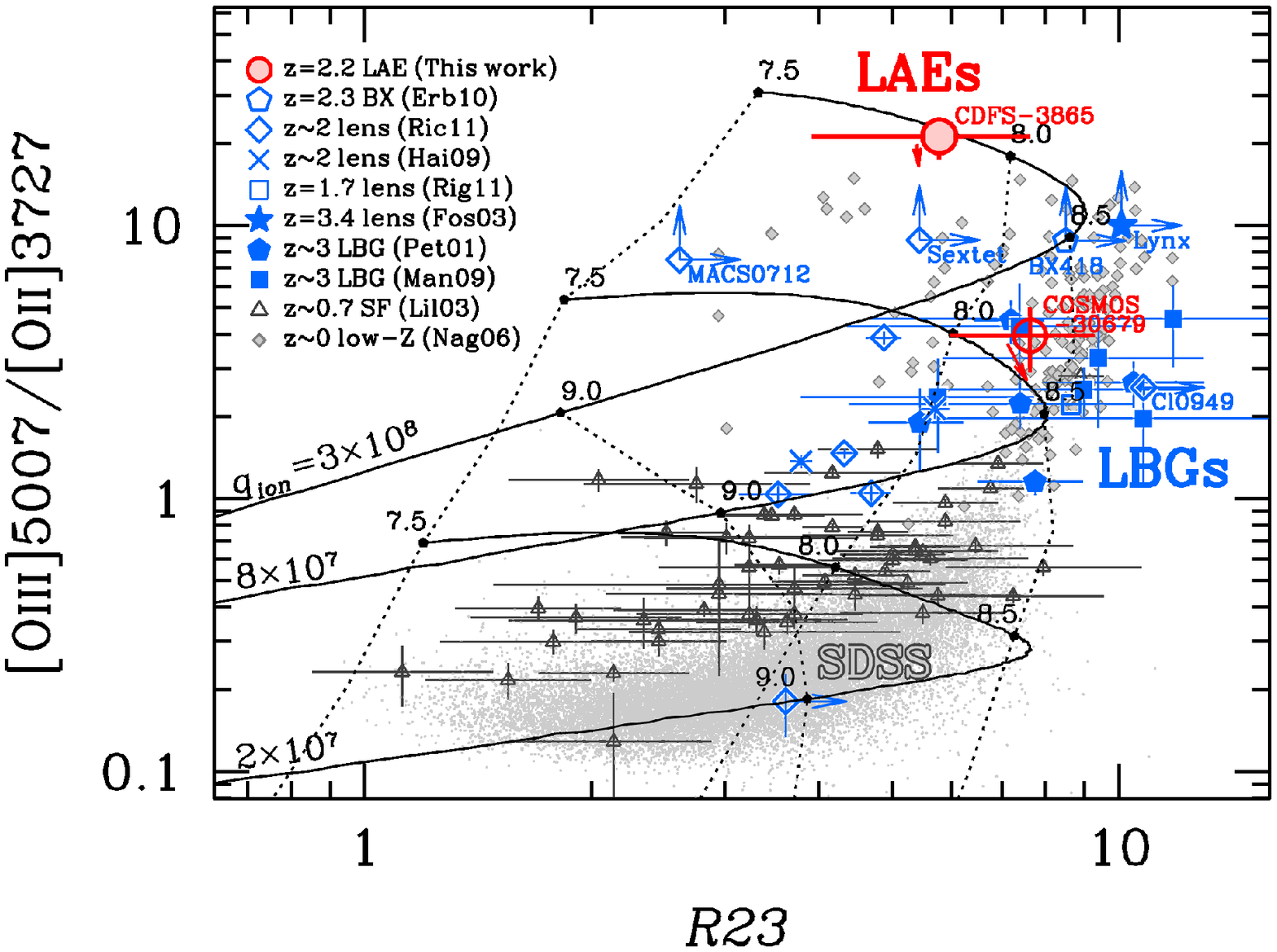}
\caption{ 
The emission line flux ratio \OIII/\OII\ vs. $R23$-index for our $z=2.2$ 
LAEs (CDFS-3865: the solid red circle, COSMOS-30679: the open red circle) 
and non-LAEs (the blue symbols); the metal-poor BX galaxy at $z=2.3$ 
(\citealt{erb2010}: the open pentagon), lensed galaxies at $z\sim 2$ 
(\citealt{hainline2009}: the crosses, \citealt{richard2011}: the open 
diamonds, \citealt{rigby2011}: the open square), the lensed 
galaxy at $z=3.4$ (\citealt{fosbury2003}: the solid star), $z\sim 3$ LBGs 
(\citealt{pettini2001}: the solid pentagons, \citealt{mannucci2009}: the 
solid squares), as well as the star forming galaxies at $0.47<z<0.92$ 
(\citealt{lilly2003}: the black open triangles), and local star forming 
galaxies in the low-metallicity range ($7\lesssim$ \Oabundance\ 
$\lesssim 8.5$: the gray diamonds) and high-metallicity range (\Oabundance\ 
$\gtrsim 8.2$: the gray dots) compiled by \citet{nagao2006}. The 
values on this plot are not corrected for dust extinction, except for the 
\citet{nagao2006}'s local data points. For CDFS-3865 and COSMOS-30679, 
effects of their extinction corrections are indicated by the red small 
arrows. The black grid shows photo-ionization model calculations 
\citep{KD2002}. For reference, the relations of 
$q_{ion}=3\times 10^8$, $8\times 10^7$, and $2\times 10^7$\,cm\,s$^{-1}$ 
are shown, linked to each other at same metallicities, which are denoted 
by \Oabundance\ values with dots.
\label{fig:o3o2r23}} 
\end{figure*}

\subsection{SFR} \label{ssec:SFR}

The \Ha\ luminosity is thought to be the most reliable SFR indicator 
relative to those based on the rest-frame UV and optical spectral features. 
Indeed, it is proportional to the birth rate of massive stars as well as 
being relatively insensitive to dust extinction as compared with 
UV-continuum. We measure the SFR of the LAEs from the \Ha\ luminosity using 
the relation \citep{kennicutt1998}:
\begin{eqnarray}
{\rm SFR}\,[M_{\odot}\,{\rm yr}^{-1}]
= 7.9\times 10^{-42} L({\rm H}\alpha)\,{\rm erg}\,{\rm s}^{-1}.
\label{eq:SFR_Ha}
\end{eqnarray}
In Table \ref{tbl:properties_spec}, we summarize the derived SFRs.

\subsection{\Ha\ Equivalent Width} \label{ssec:EW_Ha}

We calculate the \Ha\ equivalent width of an object from the \Ha\ flux 
divided by its continuum flux density derived from $K$ band photometry. 
To estimate the proper continuum, however, we have to remove the 
contribution of the \Ha\ emission on the $K$ band photometry. We follow 
the prescription given in the appendix of \citet{guaita2010} (see also 
\citealt{finkelstein2011}): 
\begin{eqnarray}
f_{\nu}({\rm H}\alpha)
 = R_{T}\times\frac{F({\rm H}\alpha)}
                   {\int T_{K}(\lambda)\frac{c}{\lambda^2}\,d\lambda}, 
\label{eq:flux_line}
\end{eqnarray}
where $f_{\nu}({\rm H}\alpha)$ is the amount of flux that \Ha\ line 
contributes to the $K$ band photometry, $F({\rm H}\alpha)$ is the observed 
\Ha\ flux, $R_{T}$ is the ratio between the filter transmission at 
$\lambda_{{\rm H}\alpha}$ and the maximum of the $K$ band transmission 
function, and $T_{K}(\lambda)$ is the $K$ band filter transmission at a 
given wavelength. The flux estimated from the calculation is subtracted 
from the $K$ band photometry. The corrections span from $0.05$ 
(COSMOS-43982) to as much as $0.63$ mag (CDFS-3865). Since COSMOS-08501 
is not detected in the $K$ band, we use the $1\sigma$ lower-limit of the 
$K$ band photometry. We thus obtain an upper-limit of the continuum 
(correction $>0.26$ mag), hence a lower-limit of the EW(\Ha) for 
COSMOS-08501.
The derived EWs(\Ha) are listed in Table \ref{tbl:properties_spec}.

\section{Discussion} 
\label{sec:discussion}

In this section, we discuss the physical properties of LAEs determined by 
this study in conjunction with \citet{finkelstein2011} and 
\citet{nakajima2012}. In the first subsection, we make comparisons in terms 
of ionization parameter, metallicity, and SFR of LAEs at $z\sim 2$ as well 
as lower-redshifts, and other galaxies at similar redshifts such as LBGs. 
We then examine the implications. In the second subsection, we show the 
properties of LAEs newly found by spectroscopy based on \Lya\ and \Ha\ 
hydrogen recombination lines. We extend the discussion to understanding the 
origins of the strong \Lya\ emission observed in LAEs.

\subsection{Comparisons between LAEs and other galaxies} 
\label{ssec:comparison}

\subsubsection{Ionization State} \label{sssec:high_qion}

We find high ionization parameters for the LAEs. In this section, we examine 
this point further by comparison to other galaxies, and discuss its 
implications.

The situation is most clearly seen in Figure \ref{fig:o3o2r23}, which shows 
the \OIII/\OII\ ratio versus $R23$-index. We plot CDFS-3865 and COSMOS-30679 
(red circles), and other galaxies such as LBGs at $z=2$--$3$ (blue symbols).
The black grid represents model predictions of \OIII/\OII\ ratio and  
$R23$-index at a given metallicity and (discrete) ionization parameter 
\citep{KD2002}.
According to the calculation, CDFS-3865 is close to the 
$q_{ion}=3\times 10^8$\,cm\,s$^{-1}$ curve. This confirms its high ionization 
parameter and low oxygen abundance. The high ionization parameter is 
suggestive of CDFS-3865 showing a very hard ionizing spectrum. Therefore, 
one of the most straightforward interpretations is that CDFS-3865 is a young 
galaxy dominated by massive stars in possibly small star-forming regions 
(expected for young \HII\ regions). Its low metallicity 
($Z\sim 0.1\,Z_{\odot}$) and large EW(\Ha) ($\sim 800$\,\AA) also support 
this idea (see also \S\ref{sssec:MZR}). The idea of LAEs being less-evolved 
galaxies is consistent with previous photometric results, such as low 
stellar masses (e.g., \citealt{pirzkal2007,ono2010b}) and small sizes (e.g., 
\citealt{bond2009,malhotra2012}).

Compared to CDFS-3865, most other $z=2$--$3$ galaxies appear to have lower 
ionization parameters. Notable exceptions are objects found by 
\citet{fosbury2003}, \citet{erb2010}, and (partly) \citet{richard2011}.
\citet{fosbury2003} and \citet{erb2010} reveal from their multi-emission 
lines analysis that their galaxies possess high ionization parameters 
($q_{ion}\gtrsim 10^9$\,cm\,s$^{-1}$), low metallicities ($Z\sim 0.1\,Z_{\odot}$ 
or less), being roughly comparable to (or more or less extreme than) those 
found in CDFS-3865. Interestingly, both galaxies turn out to exhibit strong 
\Lya\ emission. This supports our suggestion that strong \Lya\ galaxies are 
in part represented by young galaxies. 
\cite{richard2011} also find two lensed galaxies (blue diamonds with upward 
arrow) possibly have high ionization parameters 
($q_{ion}\gtrsim 10^8$\,cm\,s$^{-1}$) judged from their metallicities 
determined empirically (\Oabundance\ $=8.00^{+0.44}_{-0.50}$ and 
$8.77^{+0.14}_{-0.14}$, for Sextet and MACS0712, respectively). Unfortunately, 
\Lya\ measurements are not given in \cite{richard2011}.

However, as indicated by COSMOS-30679, LAEs' ionization parameters are 
unlikely to be always very high. The ionization parameter for COSMOS-30679 
is estimated to be $q_{ion}= 8^{+10}_{-4}\times 10^7$\,cm\,s$^{-1}$, 
relatively high but comparable to those found in other high-$z$ galaxies 
(e.g., \citealt{hainline2009}) or LBGs (e.g., \citealt{mannucci2009}), 
although the error is very large. COSMOS-30679 is thus considered to be more 
evolved than CDFS-3865, suggested from its relatively higher metallicity, 
smaller EW(\Ha), older age estimated from SED fitting, and smaller ionization 
parameter.

Next, we compare our LAEs with galaxies seen in the local universe. In 
Figure \ref{fig:o3o2r23}, the gray dots indicate SDSS galaxies with 
\Oabundance\ $\gtrsim 8.2$ (\citealt{tremonti2004,KD2002}) whose 
metallicities are derived by applying photoionization models to the most 
prominent optical emission lines (\OII, \Hb, \OIII, \Ha, \NII, \SII). 
They thus represent typical local galaxies. According to \citet{dopita2006}, 
the SDSS galaxies have ionization parameters 
$q_{ion}={\rm several}\times 10^7$\,cm\,s$^{-1}$ on average. Galaxies at 
$z=2$--$3$ including the LAEs appear to have systematically higher 
ionization parameters than the SDSS galaxies. Galaxies at intermediate 
redshift ($z\sim 0.7$; \citealt{lilly2003}) are also plotted in Figure 
\ref{fig:o3o2r23}. They appear to fall between $z\gtrsim 2$ galaxies and 
the SDSS galaxies in Figure \ref{fig:o3o2r23}, showing systematically higher 
\OIII/\OII\ ratios than the SDSS galaxies. Since their metallicities are 
comparable to those of the SDSS galaxies \citep{lilly2003,tremonti2004}, 
the higher \OIII/\OII\ ratios are likely due to their somewhat higher 
ionization parameters. Combined with the findings of much higher \OIII/\OII\ 
ratios in $z=2$--$3$ galaxies, higher-$z$ galaxies tend to have higher 
ionization parameters.

Another interesting comparison is with the low-metallicity galaxies in the 
local universe. The gray diamonds in Figure \ref{fig:o3o2r23} show local 
low-metallicity galaxies with $7\lesssim$ \Oabundance\ $\lesssim 8.5$ 
\citep{nagao2006} whose metallicities are measured with the {\it direct T$_e$} 
method (e.g., \citealt{izotov2006}). Unlike typical SDSS galaxies, they 
appear in almost the same parameter space occupied by high-$z$ galaxies on 
Figure \ref{fig:o3o2r23}. Since their low-metallicities are determined by 
the {\it direct T$_e$} method and thus reliable, the high \OIII/\OII\ ratios 
can be interpreted as due to high ionization parameters. Indeed, from the 
comparison between \OIII/\OII\ ratios and photo-ionization models given in 
\citet{nagao2006}, it is inferred that local low-metallicity galaxies 
typically have higher ionization parameters ($\sim 8\times 10^7$\,cm\,s$^{-1}$, 
or higher) than more metal-rich galaxies seen in the SDSS sample. Therefore, 
these low-metallicity galaxies in the local universe appear to have similar 
metallicities and ionization parameters as high-$z$ star-forming galaxies 
including LBGs, and particularly those with extreme quantities are likely 
analogs of high-$z$ LAEs such as CDFS-3865 in terms of their high ionization 
parameters and low metallicities.

One such extreme population is star-forming luminous compact galaxies (LCGs), 
or popularly referred to as ``green pea" galaxies (GPs; 
\citealt{cardamone2009}). The GPs are found in the SDSS spectroscopic 
sample%
\footnote{
Hence some of the GPs overlap with the local low-metallicity 
galaxies plotted with the gray diamond on Figure \ref{fig:o3o2r23}.} to have 
green colors in the SDSS $gri$ composite images, caused by the very strong 
\OIII\ emission lines with EWs of $\sim 1000$\,\AA\ falling into the $r$ 
band. They show low metallicities ($Z\sim 0.2\,Z_{\odot}$; 
\citealt{amorin2010}) and high ionization parameters 
($\gtrsim 10^8$\,cm\,s$^{-1}$) inferred from their high \OIII/\OII\ ratios 
\citep{jaskot2013}, analogous to those observed in the LAEs. Their SFRs are 
also comparable to those for LAEs \citep{izotov2011}, in contrast with more 
modest SFRs seen in typical local low-metallicity galaxies (e.g., 
\citealt{lee2004}).
A similar discussion is also provided by \citet{xia2012}.

In the SDSS Date Release 7 (DR7) spectroscopic galaxy sample, $251$ GPs 
are found \citep{cardamone2009}. Searching the entire SDSS DR7 spectroscopic 
sample yields $418,429$ objects which fall in the GPs redshift range 
($z=0.112$--$0.360$; \citealt{cardamone2009}) and are classified as GALAXY. 
Therefore, the GPs occupy only $\sim 0.06$\,\% of the SDSS galaxy sample.
Based on the luminosity function determined for the SDSS galaxies 
\citep{blanton2003}, the number density is 
$\sim 1.5\times 10^{-2}$\,Mpc$^{-3}$ above a detection limit $M_r=-17.63$, 
which corresponds to $0.037\,L^{\star}$. The limit is calculated from the 
apparent magnitude limit ($r<17.77$; \citealt{tremonti2004}) for the 
spectroscopic sample and the redshift threshold ($z>0.028$; 
\citealt{nagao2006}). The number density of the GPs is thus approximately 
$9\times 10^{-6}$\,Mpc$^{-3}$, which is almost consistent with the number 
density found by \citet{cardamone2009} ($\sim 2$ GPs per deg$^2$). On the 
other hand, the number density of LAEs at $z=2.2$ is 
$\sim 1.7\times 10^{-3}$\,Mpc$^{-3}$ calculated by integrating the luminosity 
function provided by \citet{hayes2010}%
\footnote{ 
The detection limit is $L=2.8\times 10^{41}$\,\ergs\ 
($\sim 0.02\,L^{\star}$).} above $L=0.037\,L^{\star}$. 
The number density of the GPs is thus almost two order of magnitude 
smaller than that of LAEs. 
Therefore, if the GPs alone represent galaxies with high ionization 
parameter in the local universe, although it is still uncertain, the 
large difference of abundances between LAEs and the GPs suggests that 
such galaxies appear to be much more abundant at high-$z$ while rarely 
seen in the local universe.

The final noteworthy implication of the high ionization parameters found 
in high-$z$ LAEs is that such galaxies may provide additional photons 
that cause hydrogen reionization of the IGM in the early universe. 
Previous censuses of early galaxies have revealed a possible shortage of 
ionizing photons for the cosmic reionization (e.g., 
\citealt{ouchi2009,robertson2010}). \citet{ouchi2009} find that the 
universe could not be totally ionized by only galaxies at $z=7$ if there 
is no evolution of properties (e.g., escape fraction of ionizing photons
(\fesc), metallicity, dust extinction) from $z\sim 3$ to $z=7$. 
One approach to resolve the problem is to adopt larger \fesc, which may be 
the case for galaxies with higher ionization parameter. Since lower 
ionization species such as \OII\ dominate in the outer regions of an usual, 
ionization-bounded \HII\ region (e.g., \citealt{shields1990,oey1997}), a 
density-bounded \HII\ region will show a larger \OIII/\OII\ ratio. Therefore, 
an ionization parameter depends not only on the hardness of the ionization 
radiation field but on the optical depth in an \HII\ region (e.g., 
\citealt{brinchmann2008, giammanco2005}); higher ionization parameters are 
expected to be observed from optically thin, density-bounded \HII\ regions. 
E.g., \citet{giammanco2005} calculate that in the condition of density 
bounding, an \HII\ region with \fesc=$0.5$ shows the same ionization 
parameter as a region with $\sim\times 10$ greater ionizing flux without 
any photon escape. Consequently, the high ionization parameters found in the 
LAEs may be due to a low optical depth and a high \fesc, probably achieved 
by density-bounded \HII\ regions. 

\begin{figure*}
\epsscale{1.15}
\plotone{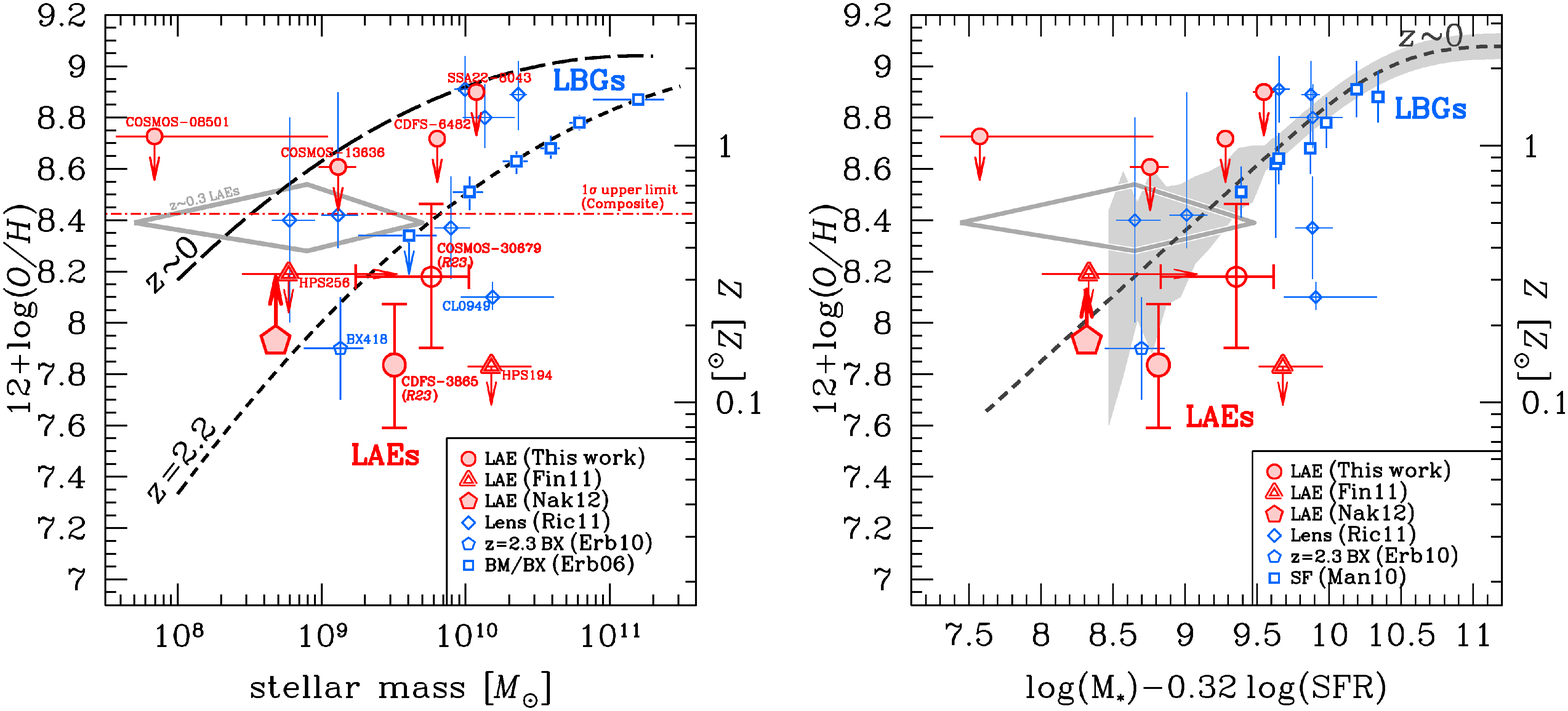}
\caption{
(left) Mass-metallicity (M-Z) relation for $z\sim 2$ LAEs (red symbols), 
lensed galaxies, and continuum selected galaxies (blue).
The circles show our LAEs, whose metallicities are measured/constrained 
by using $R23$-index (with error bar)/$N2$-index ($1\sigma$ upper-limits).
The red dot-dashed line shows the $1\sigma$ upper-limit of
metallicity ($N2$-index) for the composite spectrum in $K$ band.
We additionally plot the average LAE (pentagon: \citealt{nakajima2012}) 
and two $z\sim 2.4$ LAEs (triangles: \citealt{finkelstein2011}).
LAEs at $z=0.195-0.44$ are also shown with gray enclosed area 
\citep{cowie2011}. 
For non-LAEs, we plot lensed galaxies at $1.5<z<2.5$ (open diamonds: 
\citealt{richard2011}), the metal-poor BX galaxy at $z=2.3$ 
(open pentagon: \citealt{erb2010}), and BX/BM galaxies (open squares: 
\citealt{erb2006a}).
The M-Z relation compiled by \citet{maiolino2008} at $z\sim 0$ and 
$z\sim 2$ are drawn as long-dashed and dashed curves, respectively.
All data plotted here have been recalibrated to have the same 
metallicity scale \citep{maiolino2008} and IMF \citep{salpeter1955}
except for our LAEs with $R23$-index whose metallicities are calibrated 
with \citet{KD2002}'s relations.
(right) Fundamental metallicity relation proposed by \citet{mannucci2010}.
The dashed curve and the shaded areas indicate the relation and its 
typical dispersions respectively, determined in the local universe
\citep{mannucci2010,mannucci2011}.
The open squares show $z\sim 2$ star forming galaxies collected/compiled 
by \citet{mannucci2010} from \citet{erb2006a} and so on.
Other symbols are the same as those in the M-Z relation.
\label{fig:MZR_FMR}}
\end{figure*}

Although exact geometries of \HII\ regions in LAEs are sill uncertain, 
the LAE's optically thin property has been suggested by recent studies. 
E.g., \citet{hashimoto2013} investigate the kinematics of the galaxy scale 
outflows for LAEs using both \Lya\ emission and strong metal absorption 
lines, finding a tendency that LAEs have smaller values of neutral hydrogen 
column density in contrast with LBGs. Theoretically, \citet{yajima2012} also 
suggest a completely ionized, optically thin phase can exist for LAEs. 
Therefore, the high ionization parameters may suggest LAEs can possess 
higher \fesc, in contrast with LBGs showing modest fractions of photon 
escape. 
This trend is roughly consistent with the findings by \citet{iwata2009}
that LAEs tend to show smaller observed ratios of non-ionizing UV to 
Lyman continuum flux density than LBGs, as well as with the latest 
measurements of \fesc s for both populations \citep{nestor2013}. 
Similar suggestions of high escape fractions are also offered for LCGs, 
likely local analogs of high-$z$ emission line galaxies 
\citep{jaskot2013,shim2012}. 
Since the fraction of LAEs among star-forming galaxies is known to 
increase with redshift (e.g., \citealt{ouchi2008, stark2010}), LAEs 
and similar low-mass galaxies could produce ionizing photons efficiently 
and play a key role in supplying ionizing photons for the cosmic 
reionization at $z\gtrsim 6$.

We note in conclusion that if the photon escapes are non-zero, LAEs' 
ionizing fluxes need not necessarily to be as hard as we expect for 
zero escape fraction. A hard ionizing photon flux (young population), 
a high escape fraction of ionizing photons, or a combination of both 
conditions likely cause the high ionization parameters found in the 
LAEs.

\subsubsection{Metallicity} \label{sssec:MZR}

Figure \ref{fig:MZR_FMR} (left) shows the observed mass-metallicity (M-Z) 
relation. The red symbols represent LAEs at $z\sim 2$, while the blue symbols 
show the other galaxies such as LBGs at the similar redshifts. Although some 
of our LAEs have just weak upper-limits of metallicity due to the lack of 
\NII-detection (\S\ref{ssec:qion_Z}), for CDFS-3865 and COSMOS-30679, we 
obtain the metallicity estimates from $R23$-index with the oxygen lines.
We plot these $R23$ metallicity estimates with larger circles in Figure 
\ref{fig:MZR_FMR}. 
Albeit with its relatively large error in metallicity, CDFS-3865 falls 
below the conventional M-Z relation of $z\sim 2$ LBGs. At least one LAE 
(HPS194) found by \citet{finkelstein2011} also looks less chemically-enriched 
for its mass. Although the offsets of these LAEs from the M-Z relation can 
be due to the intrinsic scatter of the relation seen in the local universe 
(e.g., \citealt{tremonti2004}), it may also indicate that they are less 
chemically evolved for their stellar masses. The idea is consistent with 
the suggestion of CDFS-3865's high ionization parameters. Since galaxies 
with high ionization parameters (e.g., BX418) tend to fall below the M-Z 
relation as well%
\footnote{
Galaxies with high ionization parameter \citep{fosbury2003, richard2011} 
are omitted due to their high redshifts ($z>2.5$). Since the M-Z relation 
is known to evolve with redshift, a direct comparison with the $z\sim 2$ 
LAEs is not appropriate.}, young galaxies with high ionization parameters 
may not follow the M-Z relation defined by more evolved galaxies. 
Unfortunately, no constraint on ionization parameter is given for HPS194.
Alternatively, differences in star-formation activity may cause the scatter. 
We discuss this point later.

By contrast, the comparison with COSMOS-30679 is more unclear. Its large 
errors both in stellar mass and metallicity prevent us from telling if it 
is below/on the LBGs' relation. Since its ionization parameter is comparable 
to those for LBGs (\S\ref{sssec:high_qion}), COSMOS-30679 may be an LBG-like 
galaxy.

We also constrain an average metallicity from the $K$ band composite 
spectrum (Figure \ref{fig:spec_ALL_n6}) by using the empirical \NII-index.
The $1\sigma$ ($2\sigma$) upper-limit of metallicity is 
\Oabundance\ $<8.42$ ($8.66$), which corresponds to 
$Z<0.54$ $(0.93)\,Z_{\odot}$. On the other hand, we independently obtain 
an average lower-limit of metallicity for LAEs to be 
\Oabundance\ $>7.93$ ($7.63$), or $Z>0.17\,(0.09)\,Z_{\odot}$ at the 
$1\sigma$ ($2\sigma$) level, based on the \OII/(\Ha+\NII) ratio whose 
fluxes are obtained by stacking $1.18$ and $2.09\,\mu$m narrowband images 
for more than $100$ LAEs \citep{nakajima2012}. LAEs thus typically have 
a metallicity \Oabundance\ $=7.93$--$8.42$ ($7.63$--$8.66$) at the 
$1\sigma$ ($2\sigma$) level. The range is robust in the sense that the 
upper-limit is constrained by bright, massive LAEs while the lower-limit 
by faint, low-mass LAEs. More spectroscopic data are needed, however, to 
conclude that there are no exceptional LAEs that have nearly zero or 
super-solar metallicities.
The metallicity range also suggests that LAEs at $z\sim 2$ are rather less 
chemically enriched than those at $z\sim 0.3$ (\Oabundance\ $\sim 8.4$%
\footnote{
We recalculate the metallicity by using the \citet{maiolino2008} 
indicator. The original estimate is $\sim 0.15$\,dex lower.
}; \citealt{cowie2011}).

\citet{nakajima2012} find that LAEs fall typically above the M-Z relation 
below the stellar mass $\sim 10^9\,M_{\odot}$. In contrast, the current study 
finds less chemically enriched LAEs for their masses. This apparent 
inconsistency may be due to the sampling of a large variety of evolutionary 
phases within the LAE population. Such variation has been indeed reported by 
other studies (e.g., \citealt{nilsson2011,oteo2012}).
Alternatively, differences in star-formation activity may cause the 
inconsistency. To check the possibility, we plot the LAEs on the fundamental 
metallicity relation (FMR; \citealt{mannucci2010}), the relation between 
stellar mass, metallicity, and SFR. In Figure \ref{fig:MZR_FMR} (right), 
most of the galaxies at $z=0$--$2$ including the average LAE 
\citep{nakajima2012} appear to be consistent with the same FMR determined 
by the SDSS galaxies within their errors. However, some LAEs such as HPS194, 
and possibly CDFS-3865 and COSMOS-30679, still appear to fall below the 
relation. The inconsistency for HPS194 is remarkable ($\gtrsim 5\sigma$ 
level). Their low metallicities are not just likely due to their relatively 
high SFRs. We can speculate that the FMR is not universal and may fail to 
reproduce the properties of galaxies with e.g., high ionization parameter. 
Clearly however, much more data is needed to test the idea statistically.

\subsubsection{Star-Formation Activity} 
\label{sssec:MsSFRR} 

\begin{figure}
\epsscale{1.15}
\plotone{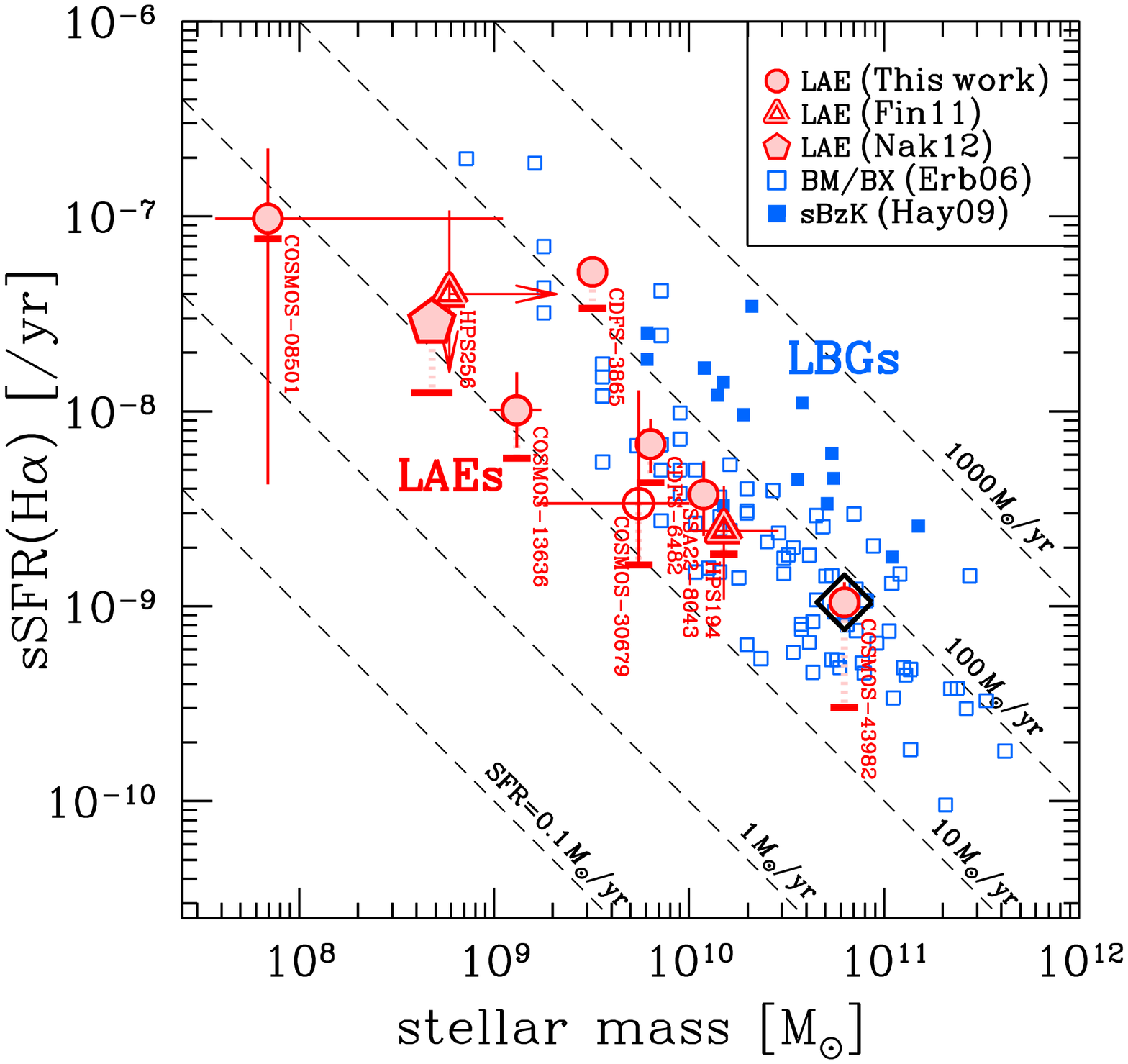}
\caption{
Relation between stellar mass and specific star formation rate (sSFR) 
for $z\sim 2$ LAEs (red) and continuum selected galaxies (blue); 
LAEs with NIR spectroscopy 
(circles: this work, triangles: \citealt{finkelstein2011}), 
the average LAE obtained by stacking NIR NB imaging 
(pentagon: \citealt{nakajima2012}), 
BX/BM galaxies (open squares: \citealt{erb2006b}), 
and sBzK galaxies (solid squares: \citealt{hayashi2009}). 
COSMOS-43982, a possible object with AGN activity, is marked with 
black diamond.
The red open circle denotes COSMOS-30679, which has less accurate 
SED fitting results due to the neighbor's contamination. 
All sSFRs plotted here are derived from \Ha\ luminosity with 
Salpeter IMF. 
The short horizontal bars in each symbols for the LAEs indicate 
the sSFR with no dust extinction correction.
The dashed lines correspond to constant SFRs of $0.1, 1.0, 10, 100,$ and 
$1000\,M_{\odot}\,{\rm yr}^{-1}$. 
\label{fig:MsSFRR}}
\end{figure}

In order to examine the star-formation activity of LAEs, we plot the LAEs 
on the sSFR versus stellar mass plane in Figure \ref{fig:MsSFRR}. We also 
plot BX/BM galaxies \citep{erb2006b} and sBzK galaxies \citep{hayashi2009}.
We note that this plot should be interpreted with cares since the 
spectroscopic data introduce limits in the sensitivity to low SFRs.

BX/BM and sBzK galaxies follow a simple scaling relation between sSFR and 
the stellar mass, whose tight relation is referred to as {\it the star 
formation main sequence} (e.g., \citealt{daddi2007}). Note that BX/BM and 
sBzK galaxies appear to have slightly different sequences; BX/BMs show 
lower sSFRs at a given mass. Compared to them, LAEs appear to follow the 
BX/BMs' main sequence almost over the full mass range. CDFS-3865 and 
COSMOS-30679, which exhibit high ionization parameters and low 
metallicities, are not outliers on this diagram.
This trend indicates that star-formation activities are well determined by 
their stellar mass, irrespective of the presence of \Lya\ emission.
Interestingly, \citet{rhoads2013} recently find that LAEs tend to have 
higher SFR surface densities for their gas mass surface densities than 
``normal'' star-forming galaxies including BzK galaxies.

\subsection{Physical Properties inferred from \Lya\ and \Ha\ emission} 
\label{ssec:Lya_Ha}

\subsubsection{Correlation between EW(\Lya) and EW(\Ha)} 
\label{sssec:EW_Lya_Ha}

\begin{figure}
\epsscale{1.15}
\plotone{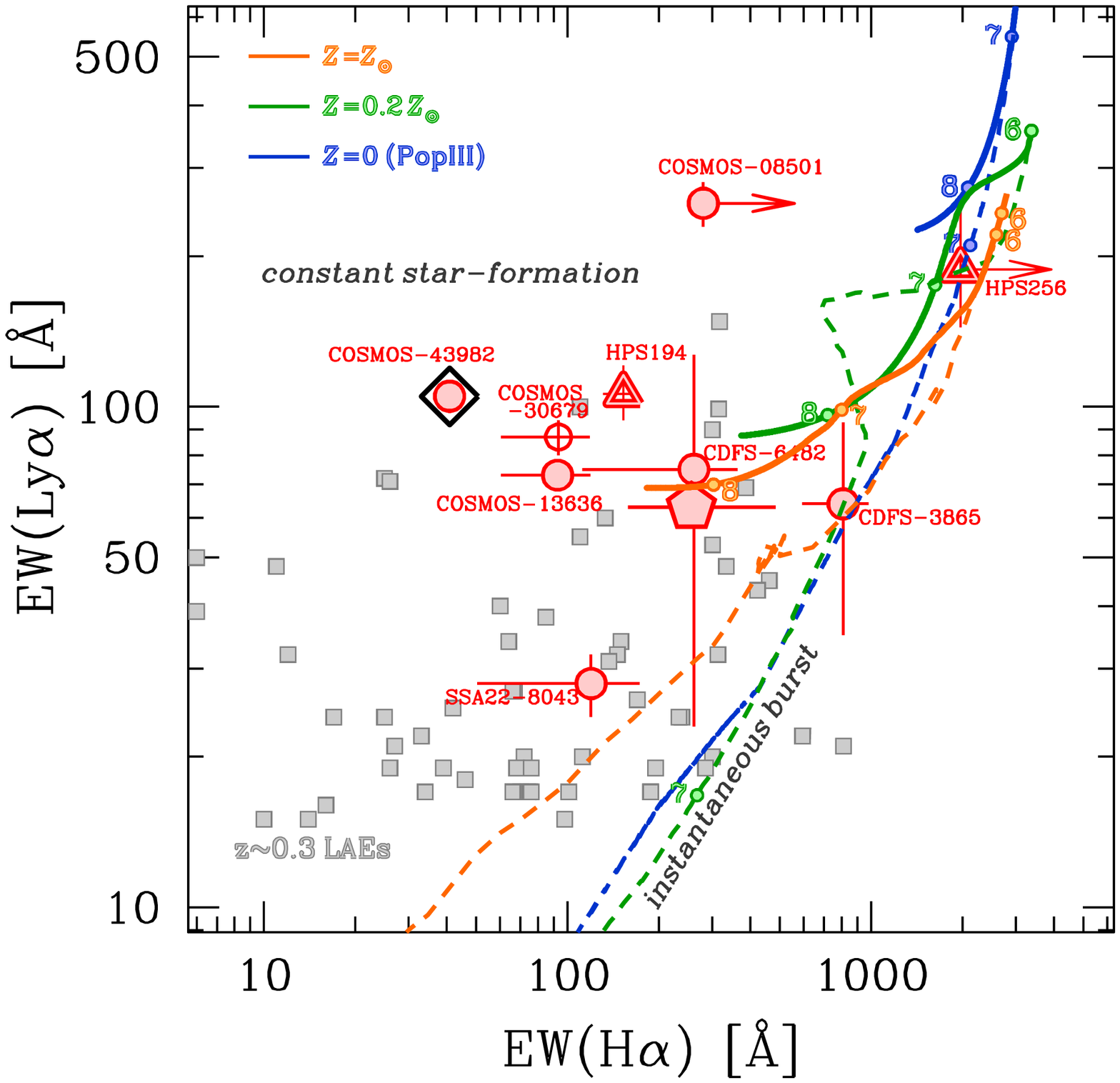}
\caption{
Relation between \Lya\ and \Ha\ equivalent widths. The red symbols show 
$z\sim 2$ LAEs (same as in Figure \ref{fig:MsSFRR}), and the gray squares 
show LAEs at $z=0.3$ \citep{cowie2011}. Superposed lines are evolutions of 
the EWs for instantaneous burst (dashed) and constant star-formation 
(solid) at metallicities of solar (orange), sub-solar (green), and zero 
(blue) calculated by \citet{schaerer2003}. Ages are denoted by the numbers 
(6, 7, 8) near the dots on the lines which indicate ($1$\,Myr, $10$\,Myr, 
$100$\,Myr). For the solar-metallicity models, the lower point labeled "6"
is for the instantaneous burst.
\label{fig:EW_Lya_Ha}}
\end{figure}

Figure \ref{fig:EW_Lya_Ha} shows equivalent widths of \Lya\ and \Ha\ for 
LAEs (hereafter referred to as the ``EWs diagram''). This plot may be useful 
to understand star-formation histories of LAEs, because the EWs' continuum 
fluxes evolve in different ways when different star-formation histories are 
assumed. Another advantage of using the EWs is that both are pure observables 
and to zeroth order independent of dust extinction (the effect of dust will 
be discussed at the end of this section).

The superposed curves on Figure \ref{fig:EW_Lya_Ha} illustrate evolutions of 
the EWs for the two extreme star-formation histories, instantaneous burst 
(dashed) and constant star-formation (solid) at several metallicities 
\citep{schaerer2003}%
\footnote{
The models are collected from the Strasbourg astronomical Data Center (CDS).
We present here the three metallicity cases assuming Salpeter IMF 
\citep{salpeter1955} with upper (lower) mass cut-off to be $100\,M_{\odot}$ 
($1$\,$M_{\odot}$). Case B recombination is assumed for an electron 
temperature of $T_e=3\times 10^4$\,K at zero metallicity and $T_e=10^4$\,K 
otherwise, and an electron density of $n_e=10^2$\,cm$^{-3}$.}. For the 
instantaneous burst, since very massive stars ($M_{\star}\gtrsim 10\,M_{\odot}$) 
complete their evolution within $\lesssim 10$\,Myr, both EWs decline rapidly. 
As a result, their curves evolve quickly to lower-left on the EWs diagram. 
For the constant star-formation, on the other hand, the EW(\Lya) stops 
declining around $\sim 100$\,Myr, because massive stars that are responsible 
for both \Lya\ emission and UV-continuum reach a steady mode. Since the 
EW(\Ha) keeps declining as the older stars build up in the galaxy, their 
slopes of the evolutionary tracks become less steep on the EWs diagram. 
Although \citeauthor{schaerer2003}'s calculations stop at $\sim 400$\,Myr
for constant star-formation, the EW(\Lya) varies little after $100$\,Myr 
\citep{CF1993}. Therefore, the tracks must extend to the left almost 
horizontally after the terminal points, reaching EW(\Ha) $\sim 100$\,\AA\ 
for the solar-metallicity case when an age of $\sim 3$\,Gyr (e.g., 
\citealt{leitherer1999})%
\footnote{
Although \citet{leitherer1999} show the EW(\Ha) evolution with age until 
$1$\,Gyr, we extend the evolution toward older age smoothly to obtain 
EW(\Ha) $\sim 100$\,\AA\ at $\sim 3$\,Gyr. We have confirmed
the validity of the smooth extrapolation for constant star-formation
by running GALAXEV \citep{BC2003}.
}.

Compared with the model tracks, some LAEs with EW(\Ha) $\sim 1000$\,\AA, 
CDFS-3865, HPS256 \citep{finkelstein2011}, and possibly COSMOS-08501, can be 
explained by the instantaneous burst models with very young ages (a few Myr).
The other LAEs whose EW(\Ha) is modest appear to prefer the constant 
star-formation models with $\gtrsim 100$\,Myr.  Although the sample is small, 
we find more than half of the LAEs appear to need a continuous star-formation 
history rather than an instantaneous burst.

Note, however, that since the EWs are sensitive to recent starbursts, LAEs 
on instantaneous tracks are not necessarily very young, but can be 
experiencing a burst after a continuous star-formation. Such a combination 
of burst plus continuous star-formation is indeed needed to explain the EWs 
of \Ha\ and optical colors observed in local dwarf galaxies \citep{lee2006}. 
An LAE with such a combined star-formation history will be on a track of 
instantaneous burst during starburst phases and on a track of constant 
star-formation for the remaining time.

\begin{figure*}
\epsscale{1.15}
\plotone{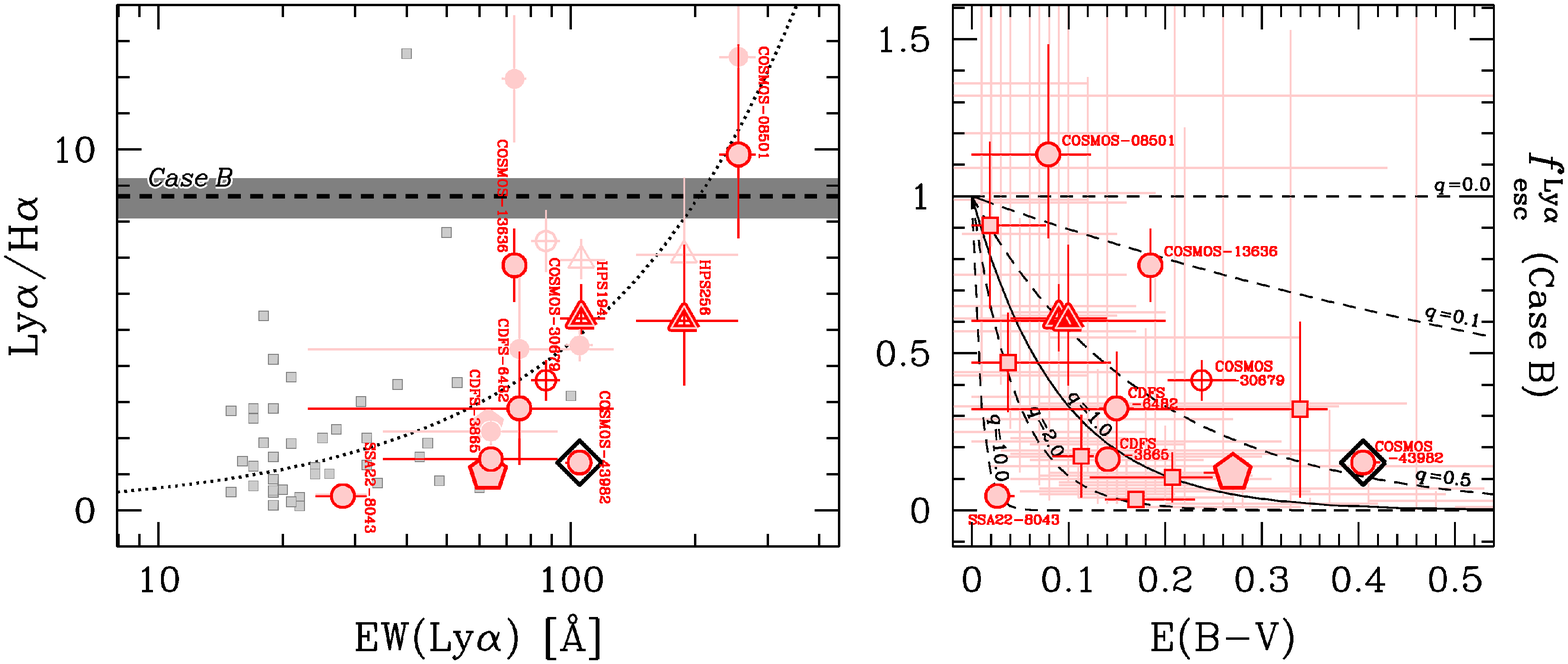}
\caption{
(left) \Lya/\Ha\ flux ratio vs. EW(\Lya). Symbols are the same as in 
Figure \ref{fig:EW_Lya_Ha}. The darker symbols show the ratios calculated 
by the observed \Lya\ luminosity divided by the intrinsic \Ha\ luminosity 
(i.e., lower-limit of the ratios), while the lighter show the ratios 
calculated by the observed \Lya\ and \Ha\ luminosities.
The ratios for $z\sim 0.3$ LAEs are based on observed luminosities. 
Dotted curve represents a single power-law fit given in Eq. 
(\ref{eq:Flux_LyaHa_EW_Lya}).
The dashed line and the gray shaded area show the \Lya/\Ha\ flux ratio 
assuming the Case B recombination (8.7; \citealt{brocklehurst1971}) and 
its variation with electron density ($8.1$--$9.2$ with 
$n_e=10^2$--$10^{10}$\,cm$^{-3}$; \citealt{HS1987}) when $T_e=10^4$\,K, 
respectively.
\%
(right) Escape fraction of \Lya\ photons (assuming the Case B) 
vs. dust extinction. 
The red squares show $z=2.2$ LAEs \citep{hayes2010}, and the faint-colored 
red error bars show $z=2$--$4$ LAEs \citep{blanc2011}. Superposed 
curves show the relations at a given 
$q$ parameter ($q=0.0, 0.1, 0.5, 1.0, 2.0, 10.0$). 
\label{fig:fesc}}
\end{figure*}

Similarly, most of $z\sim 0.3$ LAEs appear to prefer continuous 
star-formation on the EWs diagram. Interestingly, $z\sim 2$ LAEs appear to 
have systematically higher EW(\Lya) and EW(\Ha) than lower-$z$ LAEs. 
Although our spectroscopic sample may be biased toward larger EW(\Lya), the 
difference may be a sign that higher-$z$ LAEs are younger. This idea is 
supported by the inference of lower metallicities at higher-$z$ 
(\S\ref{sssec:MZR}; see also e.g., 
\citealt{finkelstein2009,cowie2010,cowie2011}).

However, one worry on the use of the EWs diagram is the effect of dust on 
\Lya. If the degree of dust extinction is different for \Lya\ and 
UV-continuum, EW(\Lya) is no longer independent of dust extinction. In order 
to examine this effect, we introduce a $q$ parameter following 
\citet{finkelstein2008}. The $q$ parameter is defined as 
$q=\tau({\rm Ly}\alpha)/\tau_{1216}$, where $\tau({\rm Ly}\alpha)$ and 
$\tau_{1216}$ are optical depth for \Lya\ and UV-continuum at 
$\lambda=1216$\,\AA, respectively. Small $q$ values ($<1$) mean \Lya\ photons 
suffer less attenuation by dust than UV-continuum photons, while large values 
mean \Lya\ photons are more heavily attenuated. In the former (latter) case 
data points go down (up) on the EWs diagram after the corrections. 
From previous works, LAEs at $z\sim 2$ have modest $q$ values; e.g., from 
the values given in \citet{hayes2010} we calculate $q\simeq 1$--$1.5$ for 
$z=2.2$ LAEs whose \Lya\ and \Ha\ luminosities are estimated from two 
narrowbands. Similarly, \citet{nakajima2012} obtain $q=0.7\pm 0.1$.
\citet{blanc2011} obtain $q=0.99$ for $z=2$--$4$ LAEs whose intrinsic \Lya\
luminosities are inferred from UV-continuum. Therefore, LAEs are on average 
likely to show $q\sim 1$, and whose observed EWs(\Lya) to be approximately 
intrinsic. However, we find a non-negligible scatter in $q$-values around 
unity (especially toward smaller values) for the individual LAEs with \Ha\ 
measurement (\S\ref{sssec:superCaseB}; Figure \ref{fig:fesc} right). The 
LAEs with $q<1$ go down on the EWs diagram and approach the instantaneous 
burst tracks. Unfortunately, with the current large errors in $q$-values, 
we cannot clearly tell which star-formation histories are likely for LAEs. 
Future radiative transfer calculations as well as higher S/N ratio spectra 
of \Lya\ and \Ha\ (and probably \Hb) will enable full use of the EWs diagram.

\subsubsection{super Case B objects?}
\label{sssec:superCaseB}

Figure \ref{fig:fesc} (left) shows the \Lya/\Ha\ ratio against EW(\Lya) for 
LAEs. The dark red symbols indicate the observed \Lya\ luminosity divided 
by the intrinsic \Ha\ luminosity corrected using the attenuation inferred 
from the SED fitting (i.e., showing lower-limits on the $y$-axis), while the 
light red symbols show the observed \Lya/\Ha\ ratios. A trend that LAEs with 
larger EW(\Lya) have larger \Lya/\Ha\ ratio appears to be present. This trend 
itself is not so surprising, but interestingly some LAEs, COSMOS-08501, and 
possibly COSMOS-13636 and one LAE at $z\sim 0.3$, may have \Lya/\Ha\ ratios 
exceeding the Case B recombination value ($8.7$; \citealt{brocklehurst1971}), 
which we call ``super Case B''. Super Case B \Lya/\Ha\ ratios have also been 
reported for some strong LAEs in the local universe (\citealt{hayes2007}, 
\citealt{atek2008}, \citealt{otifloranes2012}, and references therein). 
The variation of the intrinsic \Lya/\Ha\ ratio (the gray shaded area) does 
not seem to be a significant issue. Although it is not obvious that super 
Case B objects are really included in our sample due to their large errors, 
it is worth discussing possible physical origins of them in case they really 
exist.

We consider the possible effect of geometry, and the kinematics of dust 
and gas in the ISM. \citet{neufeld1991} propose a clumpy, multi-phase ISM 
where gas and dust are gathered in clouds within a low-density medium. With 
such circumstances, \Lya\ photons can be scattered at the surfaces of the 
clouds due to the resonant nature, while continuum photons would penetrate 
the clouds deeply. Since dust is contained in the clouds, \Lya\ photons would 
have a much smaller chance of encountering dust than other photons. In this 
scenario, a large EW(\Lya) and \Lya/\Ha\ ratio can be observed. 
Alternatively, an outflow of the ISM can be a cause of strong \Lya\ emission 
(e.g., \citealt{kunth1998}). However, \citet{hashimoto2013} find an 
anti-correlation between EW(\Lya) and \Lya\ velocity offset for LAEs, the 
latter is considered to be positively correlated with outflow velocity. 
This anti-correlation is also supported by the observations of LBGs
(\citealt{adelberger2003}; see also \citealt{shapley2003,pettini2001}).
In 
addition, the authors measure the LAEs' outflow velocity directly from metal 
absorption lines, finding that outflow velocities for LAEs and LBGs are 
comparable. Therefore, outflows do not appear to be a major mechanism for 
producing large EW(\Lya) and \Lya/\Ha\ ratio, though the sample size of LAEs 
is still small and much more data as well as theoretical works to interpret 
the properties are needed.

In order to clarify the effects of a potentially clumpy geometry of ISM on 
\Lya/\Ha\ ratios, we plot in Figure \ref{fig:fesc} (right) the relation 
between $E(B-V)$ and the escape fraction of \Lya\ photons (\fescLya) under 
the Case B recombination assumption. We estimate \fescLya\ as
\begin{eqnarray}
f_{\rm esc}^{{\rm Ly}\alpha}
 \equiv \frac{L_{\rm obs}({\rm Ly}\alpha)}{L_{\rm int}({\rm Ly}\alpha)}
 = \frac{L_{\rm obs}({\rm Ly}\alpha)}
      {8.7 L_{\rm int}({\rm H}\alpha)},
\label{eq:Lya_f_esc}
\end{eqnarray}
where subscripts {\lq}int{\rq} and {\lq}obs{\rq} refer to the intrinsic 
and observed quantities, respectively. 
The intrinsic \Ha\ luminosities are derived from the observed \Ha\ 
fluxes, corrected for dust extinctions.
The superposed lines show the 
relations at a given $q$ parameter (\S\ref{sssec:EW_Lya_Ha}); 
\begin{eqnarray}
q = \frac{-\log \left(f^{{\rm Ly}\alpha}_{\rm esc}\right)}
         {0.4k_{1216}E(B-V)}, 
\label{eq:q}
\end{eqnarray}
where $k_{1216}$ is an extinction coefficient at $\lambda=1216$\,\AA\
(11.98; \citealt{calzetti2000}).
The clumpy geometry of ISM (or outflow) is favored by objects with 
$q=0$--$1$. From Figure \ref{fig:fesc} (right), most of the LAEs presented 
here are located in the range $q=0$--$1$; e.g., COSMOS-13636 and 
COSMOS-30679 have large \fescLya\ in spite of their moderate amounts of dust. 
SSA22-8043 has a large $q$ parameter of $\sim 10$, and can be an exception 
if its \Lya\ is heavily resonant-scatted by neutral hydrogen gas.

A notable object is COSMOS-08501. It appears to fall above the $q=0$ line, 
where the clumpy ISM model does not work assuming Case B recombination, 
although the errors are relatively large. Moreover, since it is inferred to 
possess relatively small amount of dust, the large EW(\Lya) owing to the 
clumpy ISM is unlikely. In case the object is really super Case B, the 
only remaining explanation is the \Lya\ enhancement caused by collisional 
excitations. Due to the decreasing collisional strengths with increasing 
principle quantum number, collisional excitations can lead to \Lya/\Ha\ 
ratios over the Case B value (see also \citealt{osterbrock1989}). 
Shocks caused by interactions with other sources, AGN activity, supernova 
explosions, strong outflows or infall are possible candidates for the 
collisional excitations.

Based on the HST images (Figure \ref{fig:acs_I}), COSMOS-08501, a super 
Case B candidate, looks very compact and shows no sign of interactions. 
COSMOS-13636, which has a very small $q$ parameter, shows two faint sources 
nearby within $\sim 5$\,kpc (projected) from the object. Its strong \Lya\ 
emission can be (partly) due to shocks caused by interactions. Indeed, some 
fractions of LAEs are turned out to exhibit morphologies suggestive of 
mergers at higher redshift (e.g., \citealt{pirzkal2007,bond2009}) as well as 
lower redshift \citep{cowie2010}. Theoretically, \citet{tilvi2011} 
demonstrate that mergers play an important role in mass assembly and 
star-formation in majority of the LAEs especially at higher redshift. 
Unfortunately, the shock-induced \Lya\ emission due to mergings has not yet 
been taken into account in these theoretical works (see also 
\citealt{tilvi2009}).
Alternatively, as discussed by \citet{mori2004}, supernova explosions 
can cause strong shocks, resulting in strong \Lya\ emission. Although the 
authors intend to explain extended \Lya\ blobs ($\sim 100$\,kpc) with high 
\Lya\ luminosities ($\sim 10^{43}$\ergs), their basic ideas can be applied 
to normal LAEs. AGN activity and outflows seem unlikely (see \S\ref{ssec:AGN} 
and \citealt{hashimoto2013}).

When shocks (even partly) contribute to emission lines of LAEs, estimates 
of physical quantities such as ionization parameter (from \OIII/\OII\ ratio), 
metallicity (from $N2$-index), SFR (from \Ha), and dust extinction (from 
Balmer decrement) become less accurate. In particular, the intrinsic \Ha/\Hb\ 
value becomes larger when shocks are present, resulting in an overestimate 
of the abundance of dust. This effect may also help explain the presence of 
super Case B objects (e.g., \citealt{otifloranes2012}).
Thus, the results presented in this paper may require some corrections for 
the presence of shocks. We plan to address this in future work, through the 
simultaneous application of photo-ionization and shock models, with deeper
spectroscopy which will detect the weaker lines not fully detected in the 
observations presented here (e.g., Balmer lines, \OII, \OIII, \NII, \SII).

A final remark is that from our discussions so far it is evident that \Lya\ 
is not a robust indicator of SFR for LAEs. The observed data ($z=0$--$2$) 
in Figure \ref{fig:fesc} (left) are relatively well represented by a single 
power-law fit 
\begin{eqnarray}
\log \left({\rm Ly}\alpha/{\rm H}\alpha\right)_{\rm obs} 
 = && (-1.08\pm 0.29) \nonumber \\ 
   && + (0.87\pm 0.19)\times \log {\rm EW}({\rm Ly}\alpha), 
\label{eq:Flux_LyaHa_EW_Lya}
\end{eqnarray}
which is shown by the dotted curve in Figure \ref{fig:fesc} (left). 
According to this simple relation, SFRs of LAEs with EW(\Lya) $\sim 20$\,\AA\ 
(a typical threshold in narrowband searches) can be underestimated by a 
factor of about $10$. Thus, SFRs estimated from \Lya\ may involve a factor 
of $\sim 10$ uncertainties intrinsically.

\section{Summary} \label{sec:summary}

We have presented NIRSPEC and MMIRS rest-frame optical spectra of seven 
\Lya\ emitters (LAEs) at $z=2.2$, which are selected from our 
Subaru/Suprime-Cam NB387 survey in COSMOS, Chandra Deep Field South, 
and SSA22. Our first NIR spectroscopic result discusses the kinematics of 
LAEs and is presented in \citet{hashimoto2013}. As a companion study, 
this paper presents mainly the ionization and chemical properties of LAEs 
based on multiple nebular lines. Our sample includes one possibly 
AGN-dominated galaxy, and six star-forming galaxies. 
\Ha\ is detected in all six star-forming LAEs, while \NII$\lambda6584$ is 
only detected in the galaxy with signs of AGN activity. Among the six
star-forming galaxies, one (CDFS-3865) also has detections of 
\OII$\lambda3727$, \Hb, and \OIII$\lambda\lambda 5007, 4959$, and another 
(COSMOS-30679) has detections of \OII\ and \OIII. Our deep $J$ band 
spectroscopic observations provide the first \OII-detections for two 
individual LAEs at high-$z$. Our main results are summarized as follows.

\begin{itemize}

\item %
The \OIII/\OII\ ratio vs. $R23$-index diagram reveals that CDFS-3865 
has a very high ionization parameter 
($q_{ion}=2.5^{+1.7}_{-0.8} \times 10^8$\,cm\,s$^{-1}$) and a 
low oxygen abundance (metallicity; \Oabundance\ $=7.84^{+0.24}_{-0.25}$) 
in contrast with moderate values of other high-$z$ galaxies such as LBGs.
COSMOS-30679 also has a relatively high ionization parameter 
($q_{ion}=8^{+10}_{-4}\times 10^7$\,cm\,s$^{-1}$) and a low metallicity
(\Oabundance\ $=8.18^{+0.28}_{-0.28}$). LAEs would therefore 
1) represent an early stage of galaxy formation dominated by massive 
stars in compact star-forming regions, and/or 
2) have a higher escape fraction of ionizing photons probably achieved 
by density-bounded \HII\ regions. 
High-$q_{ion}$ galaxies like LAEs would thus play a key role in supplying
ionizing photons for cosmic reionization in the early universe.

\item %
Local low-metallicity galaxies ($7\lesssim$ \Oabundance\ 
$\lesssim 8.5$) show similar ionization parameters and metallicities 
to high-$z$ star-forming galaxies, and those with extreme quantities 
are likely analogs of high-$z$ LAEs. One such population is 
``green pea'' galaxies (GPs; \citealt{cardamone2009}), in terms of 
their low-metallicity, high \OIII/\OII\ ratios, and relatively high 
SFRs. A notable difference between the GPs and LAEs are their 
abundances; the GPs occupy only $\sim 0.06$\,\% of the SDSS galaxy 
sample, and its number density is almost two order of magnitude 
smaller than that of LAEs at $z\sim 2$.

\item %
CDFS-3865 falls below the mass-metallicity relation of LBGs at similar 
redshifts. Its low metallicity seems not to be explained by its star 
formation rate being taken into account, albeit with its relatively 
large error. COSMOS-30679 appears to exhibit the same trend, although 
its large error in metallicity complicates the interpretation. 
Interestingly, galaxies with high ionization parameters tend to fall 
below the relation. Such galaxies may not follow the relation 
determined by more evolved galaxies.

\item %
The composite spectrum independently provides an upper-limit on the
metallicity of \Oabundance\ $<8.42$ ($<8.66$) at the $1\sigma$ 
($2\sigma$) level. Combined with an lower-limit of metallicity 
\citep{nakajima2012}, LAEs typically have metallicities
\Oabundance\ $=7.93$--$8.42$ ($7.63$--$8.66$) at the $1\sigma$ 
($2\sigma$) level.

\item %
In contrast to the large differences in ionization parameters and 
metallicity between LAEs and LBGs, we find LAEs have similar specific 
star formation rates as BX/BM galaxies at a given stellar mass.

\item %
The EW(\Lya) vs. EW(\Ha) diagram interestingly suggests that more than 
half of the LAEs appear to need an extended star-formation such as a 
burst superimposed upon a continuous star-formation rather than the 
instantaneous burst alone. However, since EW(\Lya) may suffer from 
effects of dust and we do find a non-negligible scatter in $q$-values 
around unity (especially toward smaller values), we need to carefully 
interpret the result.

\item %
LAEs with low $q$-values ($q=0$--$1$) can be explained by the clumpy 
geometry of ISM. Interestingly, our sample may include objects with 
further enhanced \Lya, which we call super Case B. If they really exist, 
the only possible explanation is the collisional excitations of \Lya. 
Interactions with other sources and/or supernova explosions are possible 
key events that may cause shock-induced collisional excitation.
If such shocks play a role in enhancement of the \Lya\ flux, physical 
quantities such as ionization parameter, metallicity, SFR, and dust 
extinction should be re-computed using a combination of photo-ionization 
and shock-excitation models. We plan to investigate the role of shocks 
further in future works.

\end{itemize}

\acknowledgments
We are grateful to the staff of the W. M. Keck Observatory and Subaru 
telescope who keep the instruments and telescopes running effectively. 
Without their generous support, most of the observations presented here 
would not have been possible.
We thank the referee, James E. Rhoads, for his many helpful comments 
and suggestions which greatly improved this paper. 
We also thank Tomohiro Yoshikawa and Masao Hayashi for their assistance 
with NIR spectroscopy data reduction, 
Tohru Nagao and Roberto Maiolino for providing the data of emission 
lines for local galaxies, Tomoki Hayashino for providing the imaging data 
in the SSA22 field, Matthew Hayes, Eros Vanzella, and Matthew Schenker 
for their helpful comments. 
The New\Ha\ Survey team is thanked for providing their $J$ band image 
of SSA22.
This work was supported by World Premier International Research Center 
Initiative (WPI Initiative), MEXT, Japan, and KAKENHI (23244025) 
Grant-in-Aid for Scientific Research (A) through Japan Society for the 
Promotion of Science (JSPS).
KN acknowledges support from the JSPS through JSPS research fellowships 
for Young Scientists.

{\it Facilities:} %
\facility{Keck II (NIRSPEC)}, %
\facility{Magellan:Clay (MMIRS)}, %
\facility{Subaru (Suprime-Cam)}



\end{document}